\documentclass[fleqn,usenatbib]{mnras}
\usepackage{newtxtext,newtxmath}
\usepackage[T1]{fontenc}
\usepackage{xcolor}

\DeclareRobustCommand{\VAN}[3]{#2}
\let\VANthebibliography\thebibliography
\def\thebibliography{\DeclareRobustCommand{\VAN}[3]{##3}\VANthebibliography}

\usepackage{graphicx}
\usepackage{amsmath}

\title{Inferring neutron star properties through gravitational waves from r-modes and their relativistic counterparts}

\author[D. Annamalai et al.]{Dhanvarsh Annamalai,$^1$
and Rana Nandi,$^2$\thanks{E-mail:rananandi@iitd.ac.in}
\\
 $^1$Department of Physics, Shiv Nadar Institution of Eminence, Greater Noida, Uttar Pradesh-201314, India\\
 $^2$Department of Physics, Indian Institute of Technology Delhi, New Delhi 110016, India}

\date{Accepted XXX. Received YYY; in original form ZZZ}

\pubyear{2025}

\begin{document}
\label{firstpage}
\pagerange{\pageref{firstpage}--\pageref{lastpage}}
\maketitle

\begin{abstract}
         We present two frameworks to infer some of the properties of neutron stars from their electromagnetic radiation and the emission of continuous gravitational waves due to r-modes and their relativistic counterparts, termed axial-led hybrid modes. In the first framework, assuming a distance measurement via electromagnetic observations, we infer three neutron star properties: the moment of inertia, a parameter related to the mode's saturation amplitude, and the component of magnetic dipole moment perpendicular to the rotation axis. Unlike signals from mountains, axial-led hybrid oscillations provide additional information through a parameter ($\kappa$) that satisfies a universal relation with the star’s compactness. In the second framework, we utilize this and the relation between the moment of inertia and compactness, in addition to assuming an equation of state and utilizing pulsar frequency measurements, to directly measure the neutron star’s distance, along with the parameters above. We employ a Fisher information matrix-based approach for quantitative error estimation in both frameworks. We find that the error in the distance measurement dominates the errors in the first framework for any reasonable observation time. In contrast, the second framework enables accurate parameter inference because it does not depend on electromagnetic distance measurements. Although its applicability is limited to a restricted parameter space and relies on assumptions about the equation of state, the simulated errors in this framework are found to be independent of the equation of state. Finally, we discuss the potential utility and critical limitations of our analyses, and propose possible solutions and directions for future research.
\end{abstract}


\section{Introduction}

The new age of gravitational wave astronomy has the potential to provide more information than traditional astronomy or complement it in a useful way. Ever since the first detection of gravitational waves from a binary black hole merger event, GW150914, in 2015 \citep{LIGOScientific:2016aoc}, a plethora of compact binary inspirals has been detected \citep{LIGOScientific:2018mvr, LIGOScientific:2020ibl, LIGOScientific:2020aai}. These detections have given us vital astrophysical insights such as confirming the existence of stellar-mass black holes and providing an association between binary neutron star mergers, gamma-ray bursts, kilonovae, and the production of heavy elements in the universe \citep{Multimess_BNS}, constraining the equation of state high-density nuclear matter \citep{Nandi:2017rhy, Nandi:2018ami, Biswas:2020puz}, etc.

Continuous gravitational waves (CGW), as yet an undiscovered type of gravitational wave, are weak pseudo-monochromatic signals that last over long time scales, produced by tiny mountains or velocity perturbations in the star. An important potential source for CGW is r-mode oscillations in neutron stars. These are quasi-normal modes that are purely axial in Newtonian stars. However, rotating relativistic barotropic stars (where the star and its perturabations are described by the same one-parameter equation of state) do not support any pure r-modes. Since they acquire relativistic corrections with both axial and polar parity, they are referred to as axial-led hybrid (ALH) modes \citep{Lock2000,Lockitch:2002}. Both r-modes and their relativistic counterparts i.e. ALH modes are driven unstable by gravitational radiation via the CFS instability \citep{CFS, Andersson:2000mf,Lockitch:2002} with a frequency proportional to the rotation frequency of the star.
While we expect the modes to grow due to the instability \citep{Owen:1998xg} and eventually saturate because of damping, there are still uncertainties associated with the amplitude achievable by these modes because of the non-linear hydrodynamics related to the damping mechanisms that limit the growth of the modes \citep{Bondarescu:2007jw, Bondarescu:2008qx, Alford:2012yn}.  We expect that the saturation phase will last a long time, consequently generating CGW.

Although CGW signals have not been detected yet, the prospects for future detection continue to improve with the increase in sensitivity of the upcoming detectors and the refinement of the data analysis techniques \citep{Riles:2022wwz}.  Current CGW searches have only set upper limits on the saturation amplitude $\alpha_s$, a parameter that characterises r-modes during the saturation phase, from all-sky \citep{all-sky_search}, directed  \citep{direc_r-mode_1} and narrow-band searches \citep{r-mode_search_1, r-mode_search_2}.

As the prospects of detection improve, it becomes important to address what can be learnt from such a detection.  Under the assumption that a pulsar spins down just due to CGW,  a recent study \citep{Sieniawska:2021adr} showed that one can only infer the ratio of a macroscopic parameter (e.g. moment of Inertia, ellipticity) and the distance to the star. Thereafter, another study \citep{Lu:2022oys}, explored inferring the properties of the star assuming the existence of an electromagnetic distance measurement, and, that the star spins down due to a dipolar magnetic field and CGW. They also assumed that the detected signals are produced by mountains in the neutron star. This can be known with certainty, only in the case of targeted searches (when the rotation frequency of the star is known)\citep{doi:10.1142/9789811220944_0006}. The potential source could also be r-modes or other exotic possibilities \citep{Antonio:2018,Miller:2022} for candidates detected via directed or all-sky searches. It has also been shown by \citep{Ghosh:2023vja} that one can directly measure the distance to the neutron star for signals produced by r-modes when the rotation frequency of the star is known, under the assumption that the star's spindown is due only to GW emission. Based on these works, we investigate the parameter inference of a detected CGW signal produced by r-modes or ALH modes. In particular, we explore two different inference frameworks, assuming that the star spins down due to a dipolar magnetic field and CGW. In the first framework, similar to \citep{Lu:2022oys}, we assume the existence of an electromagnetic distance measurement to the star following that, we infer three neutron star properties: its moment of inertia ($I_{zz}$), the component of magnetic dipole moment perpendicular to the rotation axis  ($m_p$) and a parameter alpha ($\tilde{\alpha}$) which is related to the saturation amplitude see Section \ref{Preliminaries}). In the second framework, we explore the case of signals detected particularly via narrow-band searches, where the distance to the star can be directly measured from the CGW signal along with the other macroscopic properties mentioned before.

This paper is organised as follows. Section \ref{Preliminaries} introduces the basics of CGW and their detection. Section \ref{Parameter Estimation Framework} provides the two different theoretical frameworks used to infer neutron star properties. In Section \ref{asum}, we discuss key assumptions in our analysis. In Section \ref{Monte Carlo Simulation}, we discuss the Monte Carlo simulation used to present an error estimation study on the inferred parameters for both frameworks. Section \ref{Results} presents the results of these simulations. Section \ref{dis} provides a summary of the work, some drawbacks and potential ways to overcome them.

\section{Preliminaries} \label{Preliminaries}

In this section, we present the preliminary information relevant to this work. We discuss CFS-instability in section \ref{CFS}, the signal model of a continuous gravitational wave in section \ref{CGWSM} and the parameter estimation of the phase and amplitude parameters in section \ref{GW_parameter_estimation}.

\subsection{r-mode Instability}
\label{CFS}
In this section, we provide a brief overview of the Chandrasekhar-Friedman-Schutz (CFS) instability, following \citet{CFS}. We discuss its relevance to r-modes and summarize the behavior of these modes when analyzed within the framework of general relativity. In the Lagrangian perturbation formalism, the perturbed Euler equation is given by:
\begin{equation}
\label{lag_per}
    A_j^i\,\partial_{t}^{2}\xi^j + B_j^i\,\partial_{t}\xi^j + C_j^i\, \xi^j  = 0\, ,
\end{equation}
where $\xi^j$ is the Lagrangian displacement vector of the fluid elements and, $A_j^i, B_j^i$ and $C_j^i$ are operators containing spatial derivative ($\nabla_i$) and background stellar parameters. Readers are referred to Refs. \citet{CFS, 2019gwa} for their expressions. Assuming $\xi^j$ satisfies Eq. \eqref{lag_per}, it can be shown that the canonical energy ($E_c$) and canonical momentum ($\mathcal{J}_c$) for axisymmetric systems defined below, are conserved:
\begin{equation}
E_c=\frac{1}{2}\left[\left\langle\partial_t \xi, A \partial_t \xi\right\rangle+\langle\xi, C \xi\rangle\right] \, ,
\end{equation}
\begin{equation}
  \mathcal{J}_c =  -\operatorname{Re}\left\langle\partial_{\varphi} \xi, A \partial_t \xi+\frac{1}{2} B \xi\right\rangle
\end{equation}
Here the indices on vectors and operators are dropped. The inner product is defined as $\left\langle\eta^i, \xi_i\right\rangle=\int\left(\eta^i\right)^* \xi_i d V$. When we assume normal-mode solution to the perturabation equation of the form $\xi_j = \tilde{\xi_j}e^{i(\omega t + m\phi)}$, both these quantities are found to be related by \citep{CFS}:
\begin{equation}
\label{E,J}
E_c=-\frac{\omega}{m} \mathcal{J}_c=\sigma_p \mathcal{J}_c\, ,
\end{equation}
where $\sigma_p$ is the pattern frequency of the mode. Moreover, for normal modes, the following inequality holds:
\begin{equation}
\label{J_ineq}
\sigma_p-\Omega\left(1+\frac{1}{m}\right) \leq \frac{\mathcal{J}_c / m^2}{\langle \tilde{\xi}, \rho \tilde{\xi}\rangle} \leq \sigma_p-\Omega\left(1-\frac{1}{m}\right)\, ,
\end{equation}
where $\Omega$ is the rotation frequency of the star. 
When the system (the star) is coupled to radiation (not necessarily gravitational waves) that carries positive energy, one expects that $E_c$ is no longer conserved. Specifically, this implies $\partial_t E_c < 0$, indicating a continuous loss of energy from the system.  So any initial condition where $E_c < 0$ leads to an instability. In the rotation frequency limit ($\Omega \to 0$), co-rotating modes ($\sigma_{p} > 0$) have $\mathcal{J}_c > 0$ and counter-rotating modes ({$\sigma_p < 0$}) have $\mathcal{J}_c < 0$ (via Eq \eqref{J_ineq}). In both cases $E_c > 0$ (Eq \eqref{E,J}), which means that the two classes of modes are stable. Now for a star with finite rotation, where the pattern speed of the mode is small but positive ($\sigma_p > 0$ and $\sigma_p \to 0$), the canonical momentum is  $\mathcal{J}_c < 0$ by Eq \eqref{J_ineq}. This leads to an $E_c < 0$ (via Eq \eqref{E,J}) and therefore the onset of the instability.

This instability can be understood in the following way. Consider modes that are pro-grade (rotates along the star's angular momentum) in the inertial frame and retro-grade (rotates in the opposite direction of the star's angular momentum) in the rotating frame. The gravitational waves from these modes carry positive angular momentum from the system. Since the perturbed fluid rotates slower than the rest of the star, it has a negative angular momentum. The emission of gravitational waves makes the mode increasingly negative, leading to instability. The r-modes are particularly interesting to study, as they all satisfy the condition of being pro-grade in the inertial frame and retro-grade in the rotating frame.

The growth of the mode due to this instability is damped by viscous forces, leading to complex dynamics. Most importantly, the growth due to the instability overcomes the viscous forces only in the instability window \citep{Andersson:2000mf}, a region in the angular velocity ($\Omega$) vs temperature ($T$) plane. Key aspects of the evolution of r-modes when present within the instability window are captured by the simple phenomenological model given by Owen {\it et al} \citep{Owen:1998xg}. In this model one considers the star as a simple system with only two degrees of freedom: the uniform angular velocity ($\Omega$) and the r-mode amplitude parameter ($\alpha$), a dimensionless parameter that is proportional to the amplitude of the fluid perturbation. The crux of this model lies in the coupled evolution  of these two degrees of freedom:
\begin{equation}
\label{couple}
\begin{gathered}
\frac{d \alpha}{d t}=-\frac{\alpha}{t_{\mathrm{gw}}}-\alpha\frac{1-Q \alpha^2}{t_{\mathrm{diss}}} \\
\frac{d \Omega}{d t}=-\frac{2 Q \Omega \alpha^2}{t_{\mathrm{diss}}} .
\end{gathered}
\end{equation}
where $t_{\mathrm{gw}}$ and $t_{\mathrm{diss}}$ are the timescales associated with gravitational wave emission and viscous dissipation mechanisms, respectively. The dynamics of this model can be separated into three phases. In the beginning, the star is assumed to have a small excitation in the $l=m=2$ r-mode, either during its birth or due to accretion. Then the first phase starts, in which the r-mode grows exponentially due to Eq. \eqref{couple}. As the mode grows, it reaches the next phase where it saturates due to the non-linear hydrodynamics. This phase is modeled by the substitution of $d\alpha / dt = 0$ in Eq. \eqref{couple}. This phase is particularly relevant to our analysis, as it marks the regime where the system emits continuous gravitational waves. To facilitate discussion in the rest of the paper, we introduce the notation $\alpha_s$ to represent the value of $\alpha$ at saturation. All the equations presented in the rest of the paper are valid exclusively in this phase, where $\alpha = \alpha_s$, and no longer hold when $\alpha$ varies. In the last phase, the mode decays exponentially. Finally, we want to point out that the above discussion is within the Newtonian framework. However, the most important relativistic effect is that there are no longer any pure r-modes, rather only axial-led hybrid modes \citep{Lock2000}. In the first order in rotation frequency ($\sim \mathcal{O}(\Omega)$), the axial-led hybrid modes are r-modes plus hybrid mode corrections of the order of compactness ($\sim \mathcal{O}(C)$). Although these hybrid mode corrections slightly stabilize these modes, the conclusions that are made regarding the CFS instability on r-modes remain a good approximation even when the problem is studied in full general relativity \citep{Lockitch:2002}.

\subsection{Continuous Gravitational Wave Signal Model} \label{CGWSM}
During the saturation phase, the r-modes or their relativistic counterparts (ALH modes) produce a continuous gravitational wave (CGW) signal dominated by $l=m=2$ current quadrupole \citep{Owen:1998xg}. In this period, the noise-free strain $h(t)$  in the detector is of the form \citep{JFK-1}:
\begin{equation}
     h(t) = \Sigma_{i=1}^4 {\cal A}_i h_i(t,\vec{\lambda}),
\end{equation}
where ${\mathcal A}_i$ are functions of the amplitude parameters: the characteristic strain amplitude ($h_{0}$), the inclination angle ($\iota$) between the neutron star's rotation axis and line of sight, the polarization ($\psi$) and initial phase ($\phi_{0}$). The parameters represented by $\vec{\lambda}$ are called the phase parameters; they include the star's sky position, the gravitational wave frequency ($f$), the frequency derivatives ($f^{k}$), and if the star is in a binary system, its orbital parameters. 

The characteristic strain amplitude for r-mode oscillations is given by \citep{Owen:2010ng}:
\begin{equation}
\label{h}
    h_0 = \sqrt{\frac{512 \pi^7}{5}} \frac{G}{r c^5} f^3 \alpha_{s}MR^3\Tilde{J},
\end{equation}
where $r$ is the distance to the star from the solar system barycenter (SSB), $M$ is the mass of the star, $R$ is the radius of the star, $\alpha_{s}$ is the r-mode saturation amplitude, $f$ is the gravitational wave frequency, and $\tilde{J}$ is a dimensionless parameter that depends on the equation of state of the star and is given by \citep{Owen:1998xg}:
\begin{equation}
    \Tilde{J} = \frac{1}{MR^4} \int_{0}^{R} \varepsilon(r)\, r^6 \, dr,
\end{equation}
where $\varepsilon(r)$ is the energy density inside the neutron star. We chose to define a parameter $\tilde{\alpha} = \alpha_s MR^3\tilde{J}$, as this is a parameter that can be potentially inferred from a detection. We also highlight that the strain amplitude for r-modes is derived elegantly by recognizing that the signal emitted by a current quadrupole shares the same functional form as that produced by a mass quadrupole, with a crucial distinction: the polarization angle $\psi$ must be shifted to $\psi + \frac{\pi}{4}$. This observation allows the luminosity expression to retain its form, as it depends solely on the amplitude $h_0$. To derive Eq \eqref{h} explicitly, one computes the luminosity from the current quadrupole and equates it to the general expression (check \citep{Owen:2010ng} for details). It should be noted that although eq (\ref{h}) and the discussion following it are relevant for r-modes, we use them for ALH modes as well since the effect of general relativity is not that significant \citep{Lockitch:2002}, as discussed above.

Various mechanisms, including gravitational waves and electromagnetic radiation, could cause a neutron star to spin down. Since the observation timescale of continuous gravitational waves is much less than the intrinsic timescale of the spin-down of the star, we can model the evolution of the phase of the gravitational wave as a second-order Taylor series \citep{JFK-1, JFk-2} (here $\dot{f},\ddot{f} \equiv f^1, f^2$):
\begin{equation}
\label{4}
    \phi(t) = \phi_0 +  2\pi \left[ft + \frac{1}{2} \dot{f} t^2 + \frac{1}{6} \ddot{f} t^3\right]. 
\end{equation}
In Eq.\eqref{4},  we have ignored the detector motion w.r.t the SSB, which is a satisfactory assumption when the star's sky position is known \citep{JFk-2}, which is the case in this work. An important parameter that is relevant in the study of the spin-down of the star is the braking index $n$ given by:
 \begin{equation}
 \label{n}
     n = \frac{f \ddot{f}}{\dot{f}^2}.
\end{equation}
The value of $n$ contains information regarding the spin-down mechanism of the neutron star \citep{Riles:2022wwz}. If the star is just spinning down due to a current quadrupole moment produced by r-modes or the axial-led hybrid modes, then $n = 7$. If it is spinning down purely due to a dipolar magnetic field, then $n = 3$. If the neutron star is spinning down just due to a mass quadrupole moment (mountains in the star), then $n = 5$.

\subsection{Gravitational Wave Parameter Estimation}
\label{GW_parameter_estimation}

Assuming that a true CGW signal is not appreciably different from the signal model in section \ref{CGWSM}, we expect to estimate the ($h_0,f,\dot{f},\ddot{f}$) from the signal as they are the key parameters for inferring neutron star properties. As an initial attempt to study the errors in these parameters from a CGW detection, we use a simple Fisher information matrix approach similar to \citep{Lu:2022oys, Ghosh:2023vja, Sieniawska:2021adr}. This approach is strictly valid only in the case of high-signal-to-noise ratios. It has other issues like the possibility of a singular or ill-conditioned Fisher information matrix \citep{Fisher_issue}. Nevertheless, due to its computational simplicity, we use this approach to get a quantitative picture.  A comprehensive study using Bayesian inference is expected to give more robust results and is left to future work. 

For long observation duration compared to a day, the parameter space metric over the phase parameters ($f,\dot{f},\ddot{f}$) is approximately given by the "phase metric" \citep{Prix:2006wm}:
\begin{equation}
\label{metric}
        g_{ij}(\lambda) = \left< \frac{\partial\phi(\lambda)}{\partial f^{i}} \frac{\partial \phi(\lambda)}{\partial f^{j}}\right> - \left< \frac{\partial \phi(\lambda)}{\partial f^i} \right> \left< \frac{\partial \phi(\lambda)}{\partial f^j}   \right>\, ,
\end{equation}
where $(f,\dot{f},\ddot{f}) \equiv (f^{0},f^{1},f^{2}$) and the time average operator $\langle \quad \rangle$ is defined as:
\begin{equation}
    \langle x(t) \rangle = \frac{1}{T} \int_{-T/2}^{T/2} x(t) dt,
\end{equation}
where $T$ is the total time-span of the observation. The metric quantifies the  "distance" between two points in the parameter space. The inverse of the Fisher information matrix (which is also the covariance matrix), in terms of the metric, is:
\begin{equation}
\label{gam}
    \Gamma^{i j} = \frac{g^{i j}}{\rho^2}.
\end{equation}
In Eq \eqref{gam} $\rho$ is the signal-to-noise ratio, which for a year or longer observation time, can be averaged over $\iota$ and $\psi$ and sky position \citep{JFK-1}:
\begin{equation}
    \rho^2 = \frac{4}{25} \frac{h_{0}^{2} T}{S_h(f)}\, ,
\end{equation} Or,
\begin{equation}
\label{D}
    \rho^2 = \frac{4}{25} \frac{T}{\mathcal{D}^2}\, ,
\end{equation}
where $S_h$ is the single-sided spectral density of the strain noise in the detector, and in Eq \eqref{D}, $\mathcal{D}$ is the "sensitivity depth" \citep{D1, D2}:
\begin{equation}
    \mathcal{D} = \frac{\sqrt{S_{h}(f)}}{h_0}.
\end{equation}
The covariance matrix is calculated from Eq \eqref{gam}, using Eq \eqref{metric} and Eq \eqref{D}:
\begin{equation}
\label{f_error}
    cov(f, \dot{f}, \ddot{f})=\frac{\mathcal{D}^2}{\pi^2}\left(\begin{array}{ccc}
\frac{1875}{16 T^3} & 0 & -\frac{7875}{2 T^5} \\
0 & \frac{1125}{T^5} & 0 \\
-\frac{7875}{2 T^5} & 0 & \frac{157500}{T^7}.
\end{array}\right)
\end{equation}
Now, the only relevant amplitude parameter is $h_{0}$. The error in $h_0$ is calculated using the Fisher information matrix over the amplitude parameters $\mathcal{A}_{i}$'s and the coordinate transformation from $\mathcal{A}_{i}$ to ($h_o,\iota,\psi,\phi_0$)\citep{prix2011}. For year-long observation, the error can be averaged over the sky position and $\psi$ \citep{Lu:2022oys}:
\begin{equation}
\label{h_error}
     \sigma(h_0) \approx \frac{4.08\mathcal{D}h_0}{\sqrt{T}} \frac{\sqrt{2.59 + \cos(\iota)^2}}{1 - \cos(\iota)^2}.
\end{equation}
Note that Eq \eqref{h_error} is singular at ($\iota = 0$ or $\pi$) due to the coordinate transformation from $\mathcal{A}_{i}$ to ($h_o,\iota,\psi,\phi_0$). For this reason, we can't average over $\iota$ with a prior range that includes $0$ and $\pi$. Thus, In this work,  we assume $cos(\iota)$ lies in the range of $[-0.9,0.9]$, ignoring the small probability of cases where $|cos(\iota)| \approx 1$. 

\section{Parameter Estimation Framework}
\label{Parameter Estimation Framework}
In this section we develop two different frameworks based on \citep{Lu:2022oys} and \citep{Ghosh:2023vja}, to infer three neutron star properties: its principal moment of inertia ($I_{zz}$), the component of magnetic dipole moment perpendicular to the rotation axis ($m_p$) and a parameter ($\tilde{\alpha}$) which is related to the saturation amplitude ($\alpha_s$) by $\tilde{\alpha} = \alpha_{s} M R^3 \tilde{J}$.
In both cases, we assume that the spin-down of the star is due to magnetic dipole radiation and gravitational wave emission via r-modes or their relativistic counterparts.
We ignore the minor effects of magnetic fields of the order $10^{15}$ G or less, see section (\ref{asum}) for more details.
In the first framework, we further assume that the distance is estimated from electromagnetic observations to 20\%  accuracy, as in Ref. \citep{Lu:2022oys}. On the other hand,
the second scenario is relevant for targeted searches, where the rotational frequency of the star is known.
The frequency of an r-mode GW signal is related to the rotational frequency of the star as  $f = \ \frac {4}{3} f_{\rm rot}$. However, the frequencies of ALH-modes deviate slightly from this. They can be quantified in terms of $\kappa$ (see section \ref{Case 2}), which satisfies universal relations with compactness \citep{Idrisy:2014qca} and the dimensionless tidal deformability \citep{Ghosh:2022ysj}.  So, in the second framework, suitable only for ALH modes, we take advantage of the universal relation between $\kappa$ and compactness to infer the star's distance from CGW signals directly \citep{Ghosh:2023vja} along with the macroscopic properties mentioned above as detailed below.

\subsection{Framework 1: Distance estimated from electromagnetic observations} \label{case1}
Balancing the spin-down power with the luminosity of electromagnetic and gravitational radiation gives:
\begin{equation}
    {\frac{dE}{dt}}\Big{|}_{\rm EM} + {\frac{dE}{dt}}\Big{|}_{\rm GW} = - {\frac{dE}{dt}}\Big{|}_{\rm rot} . 
\end{equation}
The rotational kinetic energy is taken to be that of a rotating sphere:
\begin{equation}
    {\frac{dE}{dt}}\Big{|}_{\rm rot} = \frac{9}{4}\pi^2 I_{zz} f \dot{f}. 
\end{equation}
The luminosity of an electromagnetic dipole is given by:
\begin{equation}
    {\frac{dE}{dt}}\Big{|}_{\rm EM} = \frac{27\pi^4}{8c^3\mu_0} m_{p}^2 f^4,
\end{equation}
where $\mu_0$ is vacuum permeability. The gravitational wave luminosity for r-mode is \citep{Riles:2022wwz}:
\begin{equation}
\label{rmode}
    {\frac{dE}{dt}}\Big{|}_{\rm GW} = \frac{1024 \pi^9}{25} \frac{G}{c^7} \alpha_{s}^2 M^2 R^6 \tilde{J}^2 f^8 = \frac{1024 \pi^9}{25} \frac{G}{c^7} \tilde{\alpha}^2 f^8.
\end{equation}
To simplify the expressions we introduce the following constants,
\begin{equation}
    K_{d} = \frac{3\pi^2}{2 c^2 \mu_0} \quad K_{gw} = \frac{4096\pi^7 G}{225 c^7} \quad K_{h_{0}} = \sqrt{\frac{512 \pi^7}{5}} \frac{G}{c^5} .
\end{equation}
The spin-down equations are then given by:
\begin{equation}
\label{f_1}
    \dot{f} = - \frac{K_d m_{p}^{2}}{I_{zz}} f^3 - \frac{K_{gw} \tilde{\alpha}^2}{I_{zz}} f^7.
\end{equation}
Differentiating \eqref{f_1} we get:
\begin{equation}
\label{f_2}
    \ddot{f} = - 3 \frac{K_d m_{p}^{2}}{I_{zz}} f^2 \dot{f} - 7 \frac{K_{gw} \tilde{\alpha}^2}{I_{zz}} f^6 \dot{f}.
\end{equation}
As ($h_0,f,\dot{f},\ddot{f}$) are independently estimated from the CGW signal, we can solve Eq \eqref{h}, Eq \eqref{f_1}, and Eq \eqref{f_2} to get:
\begin{equation}
\label{al}
    \tilde{\alpha} = \frac{rh_0}{K_{h_{0}} f^3},
\end{equation}
\begin{equation}
\label{I}
    I_{zz} = \frac{-4K_{gw}h_{0}^2r^2f}{K_{h_{0}}^2 \dot{f} (n-3)}, 
\end{equation}
\begin{equation}
\label{m}
    m_p = \sqrt{\frac{K_{gw}h_{0}^2r^2(7-n)}{K_{d}K_{h_{0}}^2 f^2 (n-3)}}.
\end{equation}
Note that $r$ is the distance to the star and the parameter  $\tilde{\alpha}$ is independent of $n$ and is related to the saturation amplitude by:
\begin{equation}
\label{alpha}
    \tilde{\alpha} = \alpha_{s}MR^3\Tilde{J}.
\end{equation}
Here we emphasize that $\tilde{\alpha}$ is just a parameter and Eq. \eqref{alpha} is its definition. We define such a parameter as this is what one can infer from a successful detection (see Eq. \eqref{h}.) One requires mass and radius measurements from electromagnetic observations to estimate the saturation amplitude from observation. The negative sign in Eq \eqref{I} exists as $\dot{f} < 0$. Eqs \eqref{al} - \eqref{m} are valid only when $ 3 < n < 7$. This reflects the assumption that the spin-down of the star is due to magnetic dipole radiation and r-mode or ALH mode gravitational wave emission.

The differential error of any quantity is given by \citep{Lu:2022oys}:
\begin{equation}
\label{error}
    \sigma(A)^2 = \Sigma_{x,y} \frac{\partial A}{\partial x} \frac{\partial A}{\partial y} cov(x,y),
\end{equation}
where $x,y$ $\in$ ($h_0,f,\dot{f},\ddot{f}$) and "$cov$" represents the covariance of x and y. The errors of the neutron star properties can be calculated from Eq  \eqref{f_error}, \eqref{h_error}, and \eqref{error} to third order in $\frac{1}{T}$ is \citep{Lu:2022oys}:
\begin{equation}
\label{al_error}
    \frac{\sigma(\tilde{\alpha})^2}{\tilde{\alpha}^2} = \frac{\sigma(r)^{2}}{r^2} + \frac{\sigma(h_0)^2}{h_{0}^{2} }  + \frac{16875\mathcal{D}^2}{16 \pi^2 f^2 T^3},
\end{equation}

\begin{equation}
\label{I_error}
\frac{\sigma\left(I_{zz}\right)^2}{I_{zz}^2}
= \frac{4\sigma(r)^2}{r^2}  + \frac{4\sigma({h_0})^2}{h_0^2} + \frac{ 16875 \mathcal{D}^2  }{16\pi^2 f^2  (n-3)^2  T^3}
\end{equation}

\begin{equation}
\label{m_error}
\frac{\sigma(m_p)^2}{m_p^2}
= \frac{\sigma(r)^2}{r^2} + \frac{\sigma(h_0)^2}{h_0^2} + \frac{1875\,\mathcal{D}^2\left(n^2 - 12n + 21\right)^2}{16\pi^2\,f^2\,(n-3)^2(n-7)^2\, T^3}
\end{equation}

Note that the error in strain amplitude is inversely proportional to observation time ($\sigma(h_0) \sim T^{-1/2}$, see Eq \eqref{h_error}), therefore the errors asymptote to the error in distance as  $T \to \infty$:
\begin{equation}
\label{al_inf}
   \lim_{T \to \infty} \frac{\sigma(\tilde{\alpha})}{\tilde{\alpha}} = \frac{\sigma(r)}{r},
\end{equation}
\begin{equation}
\label{I_inf}
    \lim_{T \to \infty} \frac{\sigma(I_{zz})}{I_{zz}} = \frac{2\sigma(r)}{r},
\end{equation}
\begin{equation}
\label{m_inf}
    \lim_{T \to \infty} \frac{\sigma(m_{p})}{m_{p}} = \frac{\sigma(r)}{r}.
\end{equation}
The factor of 2 in Eq \eqref{I_inf} exists as $I_{zz}$ is proportional to $r^2$, whereas $\tilde{\alpha}$ and $m_p$ are proportional to $r$.

\subsection{Framework 2: Using universal relation to estimate distance} \label{Case 2}

The frequency of the relevant ($l = m = 2$) ALH-mode oscillation, under the slow-rotation approximation, is given by:
\begin{equation}
\label{kappa}
    f = |2 - \kappa| f_{rot},
\end{equation}
where $f_{rot}$ is the rotational frequency of the neutron star. For a slow-rotating Newtonian star: $\kappa = \frac{2}{3} $. It has been shown that $\kappa$ satisfies a universal relation with the compactness ($C = \frac{M}{R}$) of the star, given by (\citep{Idrisy:2014qca,Ghosh:2022ysj}):
\begin{equation}
\label{uni_1}
  \kappa = 0.667 - 0.478 C - 1.11 C^{2}. 
\end{equation}
The compactness of the star also satisfies a universal relation with the normalised moment of inertia ($\Bar{I} = \frac{I_{zz}}{M^3}$) for slowly rotating stars (\citep{Universel_1}):
\begin{eqnarray}
\label{Uni_2}
    \ln(\Bar{I}) &=& 0.8314 C^{-1} + 0.2101 C^{-2} + 3.175 \times 10^{-3} C^{-3} \nonumber \\
    & & - 2.717 \times 10^{-4} C^{-4},
\end{eqnarray}
where Eq \eqref{Uni_2} is in geometric units ($G = c = 1$). 

For the case of targeted searches, we can use Eq \eqref{kappa} to calculate $\kappa$ from a CGW detection. The universal relations can then be used to calculate the compactness and normalized moment of inertia. At this point, assuming an equation of state of the neutron star lets us calculate its moment of inertia by varying the central density to match the compactness and the normalised moment of inertia. Note that the $\bar{I}$ - $C$ relation \eqref{Uni_2} is not actually required to determine the moment of inertia as there exists a one-to-one relation between the moment of inertia and compactness for a given equation of state. We still use Eq \eqref{Uni_2} in this framework as in the case where mass measurements are available for the star, Eq \eqref{Uni_2} allows us to measure the moment of inertia independent of the EOS. We can then get the distance to the star by rearranging Eq \eqref{I}, using which the remaining parameters are calculated.  This is similar to \citep{Ghosh:2023vja}, where they use universal relations with normalized tidal deformability instead of the compactness of the star. This framework is not suitable for r-modes since there is no way to extract the moment of inertia from their GW signals, unlike ALH-modes.

The error in $\kappa$ is given by \citep{Ghosh:2023vja}:
\begin{equation}
    \sigma(\kappa)^{2} = \frac{f^2}{f_{rot}^2} \left[\frac{\sigma(f)^2}{f^2}+\frac{{\sigma(f_{rot})^2}}{{f_{rot}}^2}\right].
\end{equation}
Then using the universal relations the error in the normalised moment of inertia is:
\begin{align}
\frac{\sigma(\Bar{I})^{2}}{\Bar{I}^{2}} &= \frac{1}{C^2} \left(0.831 C^{-1} + 0.420 C^{-2} + 9.525 \times 10^{-3} C^{-3} \right. \notag \\
& \quad \left. - 1.087 \times 10^{-3} C^{-4} \right)^{2} \times \left[ \frac{\sigma(\kappa)^2}{(0.478 + 2.22 C)^{2}} + C^2 \sigma_{\kappa - C}^2 \right] \notag\\
& \quad + \sigma_{\bar{I} - C}^2
\end{align}
which is equal to the error in $\sigma(I_{zz})/I_{zz}$ given an equation of state. Here the terms $\sigma_{\kappa-C}$ and $\sigma_{\bar{I} - C}$ refer to the errors associated with the $\kappa-C$ (Eq. \ref{uni_1}) and $\bar{I}-C$ (Eq. \ref{Uni_2}) universal relations, respectively.
In our analysis, we simply model them by taking the maximum deviation of data from the corresponding fit.
This leads to $\sigma_{\kappa - C} = 2.2\%$ \citep{Ghosh:2022ysj, private} and $\sigma_{\bar{I} - C} = 9.4\%$ \citep{Universel_1}. The error in the distance of the star, calculated from Eq \eqref{I_error} is then:
\begin{equation}
\label{r_error}
   \frac{ \sigma(r)^2}{r^2} = \frac{\sigma\left(I_{zz}\right)^2}{4I_{zz}^2} - \frac{ \sigma\left(h_0\right)^2}{h_0^2}-\frac{16875 \mathcal{D}^2}{64 \pi^2 f^2(n-3)^2 T^3}. 
\end{equation}
We can also estimate the error in $\alpha$ and $m_p$ via Eq's \eqref{al_error} - \eqref{m_error}:
\begin{equation}
\label{al_error_2}
    \frac{\sigma(\tilde{\alpha})^2}{\tilde{\alpha}^2} = \frac{\sigma(I_{zz})^{2}}{4 I_{zz}^2}  + \frac{(4n^2 - 24n + 35)}{4 (n-3)^2}\frac{16875\mathcal{D}^2}{16 \pi^2 f^2 T^3},
\end{equation}

\begin{equation}
\label{m_error_2}
\begin{aligned}
\frac{\sigma\left(m_p\right)^2}{m_p^2}
&= \frac{\sigma(I_{zz})^2}{4 I_{zz}^2} + \frac{1875 \mathcal{D}^2\left(n^2-12 n+21\right)^2}{16 \pi^2 f^2(n-3)^2(n-7)^2 T^3}  \\
&\quad - \frac{16875 \mathcal{D}^2}{64 \pi^2 f^2(n-3)^2 T^3}
\end{aligned}
\end{equation}
Note that all the neutron star properties except the distance are independent of the signal strain amplitude ($h_0$).  

\section{Assumptions}
\label{asum}
In this section, we present further discussion on the key assumptions made in our analysis.  Firstly, we assume that a supposed CGW detection can be identified as a signal due to a r-mode or its relativistic counterpart. It might not be possible if the rotational frequency of the star is not known \citep{doi:10.1142/9789811220944_0006}. In that case, only the value of the braking index ($n$) can give us an idea of the spin-down mechanism. The CGW signal detected in an all-sky survey could also be due to more exotic sources \citep{Antonio:2018, Miller:2022}.  However, a successful detection with the gravitational wave frequency ($f \approx \frac{4}{3} f_{\rm rot}$) in targeted/narrow-band searches would strongly imply that the signal is produced by r-modes or their relativistic equivalents \citep{ngnk-xmgs}.

We assume that the star spins down due to magnetic dipole radiation and Gravitational wave emission via r-modes or ALH-modes. This leads to a braking index of $n \in [3,7]$, which is inconsistent with current observations \citep{n_obs}. Recent NICER measurements also provide evidence for non-dipolar magnetic fields \citep{NICER}. Alternative spin-down mechanisms and complex magnetic field models consistent with current observations are still works in progress \citep{spin-down_review}.  Numerous factors other than relativistic effects, such as the magnetic field, influence the mode frequency of these modes\citep{mag_effect_r_mode}. However, we have ignored these factors due to their negligible effects \citep{Idrisy:2014qca}. Additionally, it's worth noting that r-modes themshelves could potentially amplify the magnetic field \citep{mag_effect_r_mode_1}, a consideration which we have chosen to disregard. In this work, we also don't consider stratification, which is shown to be important for realistic mature neutron stars \citep{Gittins:2022rxs}.

 In this work, we mostly use relations that are valid for $l = m = 2$ Newtonian r-modes (Like Eq. \eqref{h} and Eq. \eqref{rmode}), except the universal relation  (Eq. (33)), which considers the relativistic corrections to the frequency of the r-modes. To be fully consistent we must also consider the post-Newtonian  corrections to those relations \citep{Lock2000}. We chose to ignore these $\sim \mathcal{O}(\frac{M}{R})$ terms, as it gives us a simple phenomenological model for inference and unlike the frequency of the r-mode \citep{Idrisy:2014qca}, there is no work done to understand if the relativistic correction to the mode amplitude is the major correction compared to other effects like the magnetic fields.

The universal relations used in the second framework are valid only in a restricted parameter space. The $\kappa - C$ relation is valid only in the slow-rotation approximation \citep{Ghosh:2022ysj} and has corrections of the order of $(\frac{f}{f_k})^2$ \citep{Idrisy:2014qca}, where $f_k$ is the Kepler frequency. The $\bar{I}-C$ relation is also not valid for rapidly rotating stars \citep{Universel_1} and must be affected by magnetic fields. Similar universal relations called the "I-Love-Q" relation \citep{Yagi:2013bca}, have been shown to become EOS dependent for magnetic fields $B > 10^{12}$G \citep{Uni_Mag}. Although to the best of our knowledge, no similar study exists for $\bar{I}-C$ relations used here, such a limit would further restrict the region where the second inference framework can be used.

It should be noted that one can estimate the moment of inertia in framework 2 using the $\kappa$–$C$ relation, provided an equation of state (EOS) is assumed, without relying on the $\bar{I}$–$C$ relations. In this approach, the aforementioned restriction on the parameter space would not pose an issue. However, we chose not to follow this route because the use of the $\bar{I}$-$C$ relations offers the possibility of inferring the moment of inertia directly and consequently all other relevant parameters, without making any assumption about the EOS.

\section{Monte Carlo Simulation} 
\label{Monte Carlo Simulation}
In this section, we explain how we use Monte Carlo simulations to study the errors in the three inferred parameters. In Section \ref{monte1}, we talk about Monte Carlo simulations for the first framework (Section \ref{case1}). Then, in Section \ref{monte2}, we discuss the procedure of Monte Carlo simulations for the second framework (Section \ref{case1}). 

\subsection{Framework 1}
\label{monte1}

Equations \eqref{al}, \eqref{I}, and \eqref{m} suggest that to infer the values of $\tilde{\alpha},I_{zz}$, and $m_p$, we need to input the values of $r,h_0,f,\dot{f}$, and $\ddot{f}$ (or equivalently $n$). Additionally, we require $\mathcal{D},T,\iota$, and $\Delta r$ in order to estimate their errors (see Eqs. \eqref{h_error}, \eqref{al_error}, \eqref{I_error}, and \eqref{m_error}).
Therefore, we simulate the signal from a neutron star by 9 parameters: $f_{\rm in}, \dot{f}_{\rm in}, \ddot{f}_{\rm in}, I_{{zz}_{\rm in}}, r_{\rm in}, T_{\rm in}, \mathcal{D}_{\rm in}, \cos(\iota)_{\rm in},$ and $ \Delta r_{\rm in}$. Since the strength of a continuous GW signal depends not only on the value of $I_{zz}$ but also on the other neutron star parameters, $I_{zz}$ does not relate directly to the signal detectability, unlike $h_0$. Hence, using $I_{zz}$ as input is preferable to $h_0$ \citep{Lu:2022oys}, and we do it by rearranging Eq \eqref{I}.
We draw the values of $I_{{zz}_{\rm in}},f_{\rm in},\dot{f}_{\rm in},n_{\rm in},$ and ${\cos(\iota)}_{\rm in}$ from uniform distributions the ranges of which are given in Table \ref{tab:input_params_1}. For $r_{\rm in}$, $\Delta r_{\rm in}$, and $ \mathcal{D}$, we choose suitable fixed values as described below, and $T_{\rm in}$ is varied between $0.5$ and $4$ years.
Then we calculate $h_{\rm in}$ via Eq \eqref{I} and use it to determine the values of $\alpha_{\rm in}$, and ${m_p}_{\rm in}$ via Eq \eqref{al} and \eqref{m}.

The errors ($\delta f,\delta \dot{f},\delta \ddot{f}$) are drawn from a multivariate normal distribution whose covariance matrix is given by Eq \eqref{f_error}. Similarly, the error ($\delta h_0$) is drawn from a normal distribution with the standard deviation given by Eq \eqref{h_error}.
All other covariances are assumed to be zero. Then the output parameters are given by:
 \begin{equation}
 \label{out}
\begin{array}{lll}
f^{\text {out }}=f^{\text {in }}+\delta f, & \dot{f}^{\text {out }}=\dot{f}^{\text {in }}+\delta \dot{f}, \\
\ddot{f}^{\text {out }}=\ddot{f}^{\text {in }}+\delta \ddot{f}, & h_0^{\text {out }}=h_0^{\text {in }}+\delta h_0 .
\end{array}
\end{equation}
Using these output parameters we calculate $I_{{zz}_{\rm out}}, \tilde{\alpha}_{\rm out},$ and $ {m_{p}}_{\rm out}$ (with the help of eqs. \eqref{al}, \eqref{I}, and \eqref{m}) which are then compared with $I_{{zz}_{\rm in}}, \tilde{\alpha}_{\rm in},$ and $ {m_{p}}_{\rm in}$ . This process is iterated $10^6$ times. 

\subsubsection{Choice of input parameters} 

\begin{table}[h]
\centering
\caption{Choice of input parameters of the Monte Carlo simulation for framework 1.}
\begin{tabular}{|l| l| } 
\hline 
 {\bf Input parameter} & {\bf Choice of parameter value/range}\\
 \hline 
 $T_{\rm in}$ &  $0.5 - 4$ years \\ 
 \hline
 $r_{\rm in}$ & $1$ Kpc \\ 
 \hline
 $\Delta r_{\rm in}$ & $20\%$\\ 
 \hline
 $\mathcal{D}_{\rm in}$ & $30$ Hz$^{\frac{-1}{2}}$\\ 
 \hline
 $I_{{zz}_{\rm in}}$ & $(1-3) \times 10^{38}$ Kg-m$^{2}$ \\ \hline 
 $\cos(\iota_{in}$) & $-0.9-0.9$ \\ 
 \hline
 $n_{\rm in}$ & $3-7$ \\ 
 \hline
 $f_{\rm in}$ & $30 - 700$ Hz\\ 
 \hline
 $\dot{f}_{\rm in}$ & $(-10^{-8})- (-10^{-12})$ Hz s$^{-1}$\\ 
 \hline
\end{tabular}\label{tab:input_params_1}
\end{table}

\label{case 1:Choice of input parameters input Parameters}
  Table \ref{tab:input_params_1} shows the choice of the ranges/values of the 9 input parameters mentioned above. We consider an observation time of $0.5-4$ years, as gravitational wave detectors observing runs last at least a year. A fixed $r_{\rm in}$ of $1$ Kpc is assumed. This is expected to be within the range where all-sky searches are sensitive to neutron stars with $\alpha_{s} > 10^{-4}$ and $f > 100$ Hz. An error of 20\% is assumed in the distance measurement, which is not unreasonable for current and next-generation radio telescopes \citep{Dis_error, d_error}. 

Unlike \citep{Lu:2022oys}, we only explore a sensitivity depth of $\mathcal{D}_{in} = 30$ Hz$^{-1/2}$, which is typical for all-sky searches \citep{D1}. This must be interpreted as a signal being detected with a relatively low computational cost. A more sensitive follow-up analysis would significantly increase the signal-to-noise ratio. Check \citep{Lu:2022oys} for details on how stability time, the time taken for the parameter errors to reach within 10\% of the distance error, varies as a function of sensitivity depth. One could also use $S_h(f)$ of current and future gravitational wave detectors \citep{Ghosh:2023vja}, but we don't take this approach. 

We draw the moment of inertia from the widely accepted range
for neutron stars of $I_{{zz}_{\rm in}} = [1,3] \times 10^{38}$ Kg-m$^{2}$ \citep{I_range_1, I_range_2, I_range_3}, ${\cos(\iota)}_{\rm in}$ is drawn from $[-0.9,0.9]$ and $n_{\rm in}$ is drawn from $[3,7]$, based on our assumption of the spin-down mechanism. The range of values assumed for $f_{\rm in} $ and $\dot{f}_{\rm in}$ is given by:
\begin{equation}
    f_{\rm in} = [30 , 700]\, \text{Hz}, \quad \dot{f}_{\rm in} = [-10^{-8},-10^{-12}]\,\text{Hz s}^{-1},
\end{equation}
where the range of $f$ is slightly smaller than current all-sky surveys \citep{f_range}, as for higher frequencies we would have to include the corrections to r-mode frequency due to rapid rotation \citep{Idrisy:2014qca}. These ranges almost translate to surface magnetic fields of the order $10^{11} - 10^{15}$ G. We also restrict saturation amplitude to values between $[10^{-7}, 10^{-1}]$, which is consistent with current numerical simulations \citep{Arras:2002dw, Bondarescu:2008qx}.

\subsubsection{Output Parameters} 
\label{Output Parameters:case 1}

As discussed above, after choosing values for the input parameters, we calculate the output parameters $h_{\rm out},f_{\rm out},\dot{f}_{\rm out}, $ and $ \ddot{f}_{\rm out}$ via Eq \eqref{out}, and consequently the output braking index $n_{\rm out}$ using Eq \eqref{n}. Then  $I_{{zz}_{\rm out}}, \tilde{\alpha}_{\rm out},$ and $ {m_{p}}_{\rm out}$ are calculated via Eqs \eqref{al}, \eqref{I}, and \eqref{m}). We convert $\tilde{\alpha}_{\rm out}$ into estimates of $\alpha_{s,{\rm out}}$ by using Eq \eqref{alpha} and fiducial values for $M = 1.4 M_\odot$, $R = (5 I_{{zz}_{\rm in}}/2 M)^{1/2}$ and $\Tilde{J} = 0.01635$ \citep{Owen:1998xg}.

For some set of input parameters (referred to as data), we obtain $n_{\rm out} \not\in [3,7]$ and/or $ \alpha_{s,{\rm out}} \not \in [10^{-8}, 10^{-1}]$. We discard such sets of data. Only a small fraction of the data ( $\sim 1\%$) is rejected for $n_{\rm out}$ falling outside the desired range for observation time greater than a year. On the other hand, we ignore around 8\% data when the value of $\alpha_{s,{\rm out}}$ falls outside the indicated range for $n_{\rm in}=3$.
However, as $n_{\rm in}$ increases from 3 to 7, more cases with $\alpha_{s,{\rm out}}> 10^{-1}$ are encountered, and we need to omit up to 16\% of data. Although a significant chunk of data is ignored in such a way, we still obtain key quantitative trends (section [\ref{Results}]).

To compare the output and input parameters we use the median relative error:
\begin{equation}
\label{median_error}
\epsilon(P) \equiv \operatorname{median}\left\{\left.\frac{\left|P_{\text{out}}-P_{\text{in}}\right|}{P_{\text{in}}} \right|\,\, \text{with}\,\, P \in \{I, m_p, \tilde{\alpha}, r\} \right\},
\end{equation}
following \citep{Lu:2022oys}. For framework 1, we also normalize the error of ($I_{zz},m_p,\alpha$) by the median error in the distance ($\epsilon(r) \approx 0.135$) \citep{Lu:2022oys}:
\begin{equation}
\label{normal}
    \tilde{\epsilon}(I_{zz}) = \frac{\epsilon(I_{zz})}{2\epsilon(r)}, \quad \tilde{\epsilon}(\tilde{\alpha}) = \frac{\epsilon(\tilde{\alpha})}{\epsilon(r)},
    \quad \tilde{\epsilon}(m_p) = \frac{\epsilon(m_p)}{\epsilon(r)}. 
\end{equation}

\subsection{Framework 2} \label{monte2}

The key parameters required for inference in this framework are $h_0,f,\dot{f},n,$ and $\kappa$ and their errors depend on $\mathcal{D}, T,\iota, h_0,\sigma_{\kappa - C}, \sigma_{\bar{I} - C}$ and $ \Delta f_{rot}$. Here, $\Delta f_{rot}$ refers to the measurement error in the neutron star's rotation frequency from electromagnetic observations. In this case, we directly input values of $h_0$, as we can't directly input values of any internal neutron star properties. The 9 input parameters are then $f_{\rm in}, \dot{f}_{\rm in}, n_{\rm in}, {h_0}_{\rm in}, \kappa_{\rm in}, T_{\rm in}, \mathcal{D}_{\rm in}, \cos(\iota)_{\rm in}$, and $ \Delta f_{rot}$.  We draw the values of $h_{0\rm in}, \kappa_{\rm in}, f_{\rm in},\dot{f}_{\rm in},n_{\rm in},$ and ${\cos(\iota)}_{\rm in}$ from uniform distributions, $\mathcal{D_{\rm in}}$, and $\Delta f_{\rm rot}$ are set to a fixed values, and $T_{\rm in}$ is taken from 0.5 to 4 years (see Table \ref{tab:input_params_2} and discussion below). Using the value of $\kappa_{\rm in}$, we first calculate $C$ by inverting Eq \eqref{uni_1} and add the corresponding error $\sigma_{\kappa - C}$. Then we calculate the normalized moment of inertia $\bar{I}$ via Eq \eqref{Uni_2} and by taking the error $\sigma_{\bar{I} - C}$ into consideration. An equation of state is then assumed to calculate $I_{{zz}_{\rm in}}$ as detailed in section \ref{Case 2} and consequently $r_{\rm in}, \tilde{\alpha}_{\rm in},$ and $ {m_{p}}_{\rm in}$ via  Eq's \eqref{al} - \eqref{m}. We finally calculate the output parameters similar to section \ref{monte1}, and this is iterated $10^5$ times (An order less than the previous case due to computational cost).
\subsubsection{Choice of input parameters }
\label{Choice of input parameters input Parameters:case2}

\begin{table}
\centering
\caption{Choice of input parameters of the Monte Carlo simulation for framework 2.}
\begin{tabular}{ | l | l| } 
  \hline
 {\bf Input parameter} & {\bf Choice of parameter value/range}\\
 \hline
 $T_{\rm in}$ &  $0.5-4$ years \\ 
 \hline
 ${h_0}_{\rm in}$ & $10^{-27}- 10^{-25}$ \\ 
 \hline
 $\kappa_{\rm in}$ & $0.45- 0.60$\\ 
 \hline
 $\mathcal{D}_{\rm in}$ & $100$ Hz$^{\frac{-1}{2}}$\\ 
 \hline
 ${\Delta f_{\rm rot}}$ & $0.1\%$ \\ 
 \hline
 cos($\iota_{\rm in}$) & $-0.9-0.9$ \\ 
 \hline
 $n_{\rm in}$ & $3-7$ \\ 
 \hline
 $f_{\rm in}$ & $30-500$ Hz\\ 
 \hline
 $\dot{f}_{\rm in}$ & $(-10^{-8})-(-10^{-12})$ Hz s$^{-1}$\\ 
 \hline
\end{tabular}
\label{tab:input_params_2}
\end{table}

Table \ref{tab:input_params_2} shows the choice of the ranges/values of the 11 input parameters mentioned above. We consider a similar observation time to the previous case. A range of $[10^{-27}-10^{-25}]$ is explored for ${h_0}_{\rm in}$, as this range includes signals that current and future detectors can potentially detect on narrow band searches \citep{Upper_lim_narrow, Riles:2022wwz}. The range of ${h_0}_{\rm in}$ can be interpreted as a range in the input distance ($r_{\rm in}$) of the star since it is the sole parameter that depends on the strain amplitude ($h_0$).  The value of  $\kappa$ is drawn from $[0.45,0.60]$, based on theoretical considerations \citep{Idrisy:2014qca}. We explore a sensitivity depth of $\mathcal{D}_{\rm in} = 100$ Hz$^{-1/2}$, which is reasonable for narrow-band searches \citep{Wette:review, D1}. Similar to the previous case, $\cos(\iota)$ is drawn from $[-0.9,0.9]$ and $n_{\rm in}$ is drawn from $[3,7]$. The range of values assumed for $f_{\rm in} $ and $\dot{f}_{\rm in}$ is given by:
\begin{equation}
    f_{\rm in} = [30 , 500]\,\, \text{Hz}, \quad \dot{f}_{\rm in} = [-10^{-8},-10^{-12}] \,\text{Hz s}^{-1},
\end{equation}
which is a smaller parameter space in $f$ in comparison to the previous case, as the universal relation (Eq \eqref{Uni_2}) is only valid for slow rotations. We estimate the rotation frequency of the star from Eq \eqref{kappa} and assume a measurement error ($\Delta f_{rot}$) of 0.1\%, although pulsar frequencies are measured with much higher accuracy \citep{pulsar_tim}. These ranges almost translate to similar values of magnetic field and saturation amplitude as found in section \ref{monte1}.

\begin{figure*}
 \centering
 \begin{tabular}{cc}
  \includegraphics[width=0.46\textwidth]{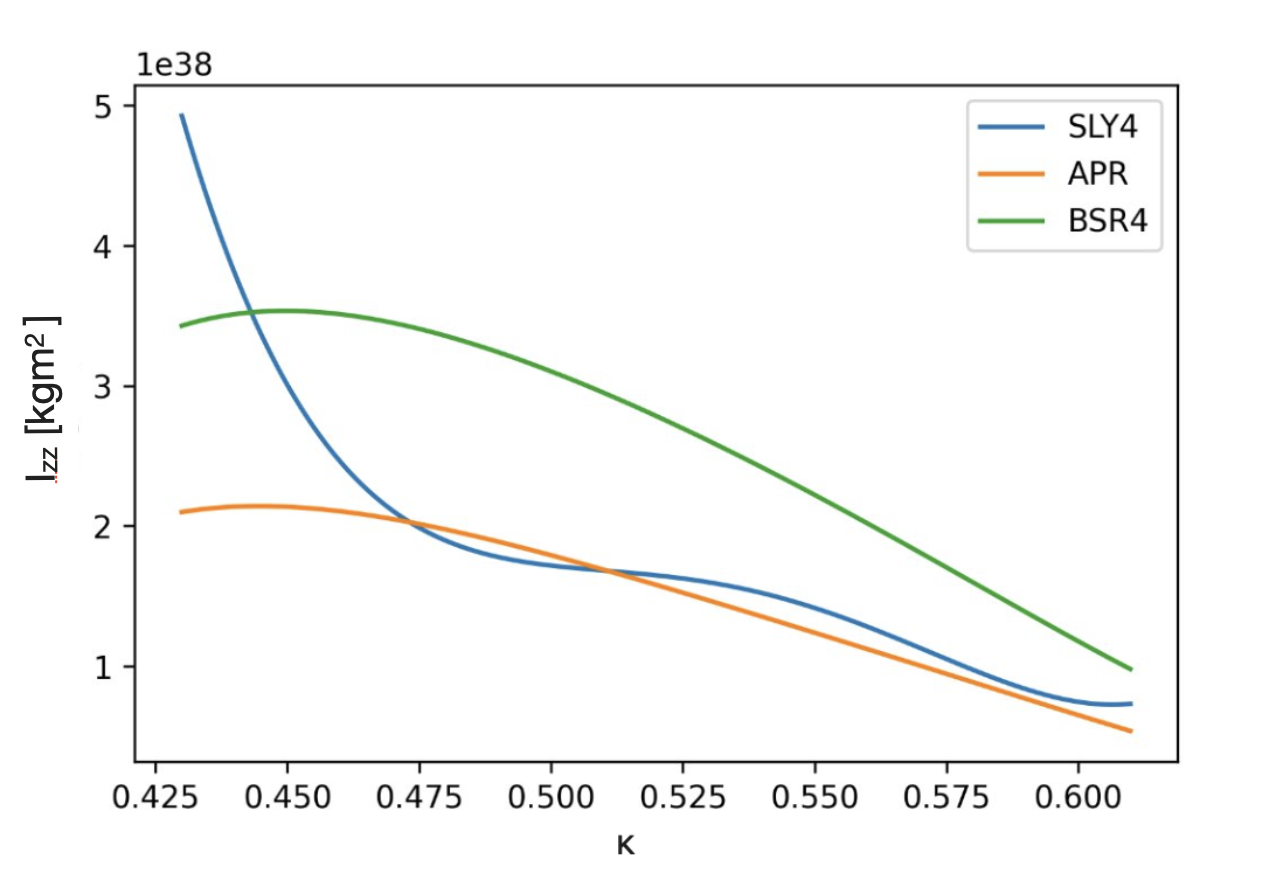} &
   \includegraphics[width=0.492\textwidth]{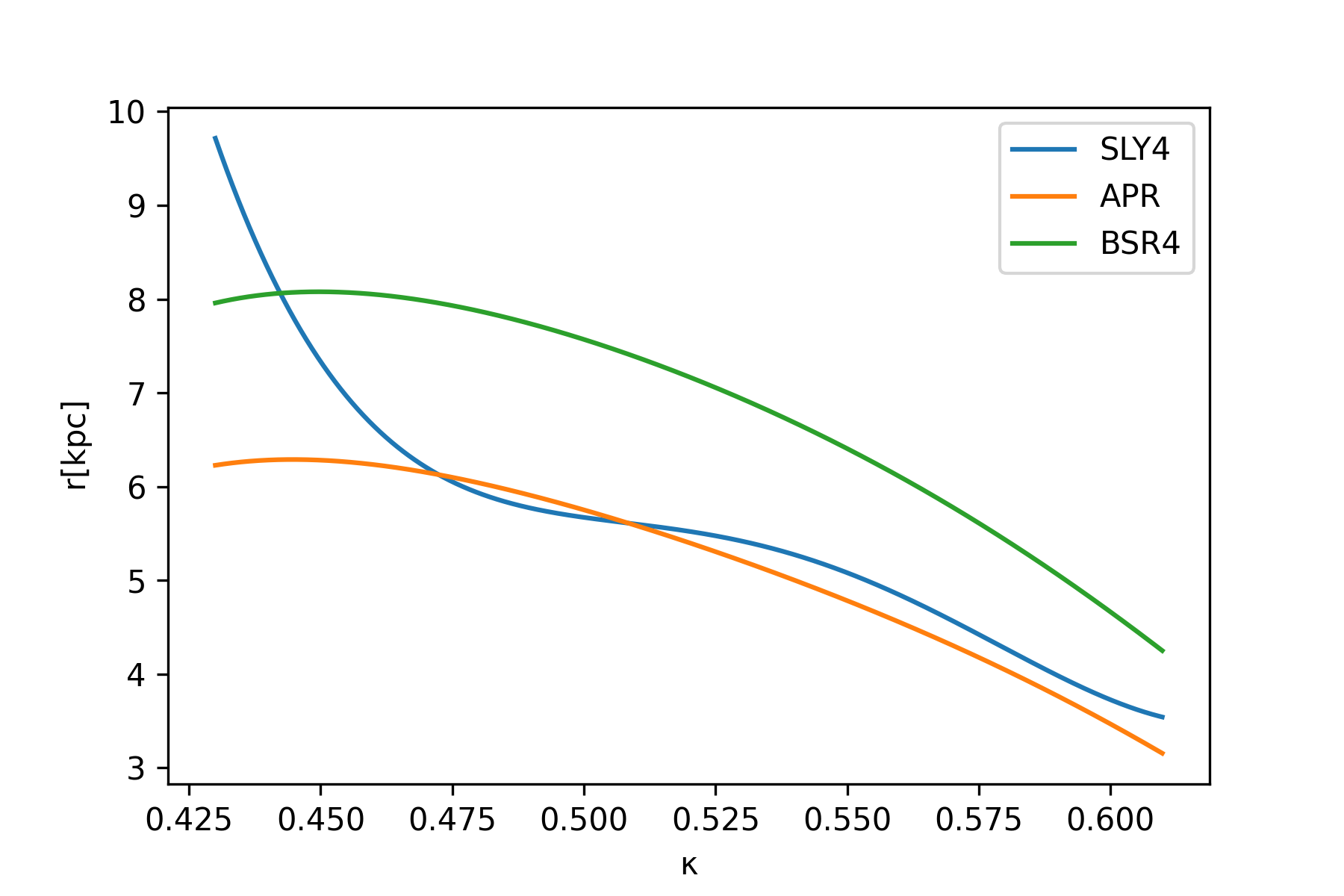}
 \end{tabular}
  \caption{(Left panel) Moment of Inertia ($I_{zz}$) as a function of $\kappa$ for various equations of state. (Right panel) Distance(r) as a function of $\kappa$, where $n = 4$, $f_{in} = 500$ Hz, $\dot{f} = -1 \times 10^{-9}$, and $h_0 = 2 \times 10^{-25}$.}
    \label{fig:9}
\end{figure*}
In this work, we use the BSR4 (\citep{BSR4_1, Nandi:2020luz}) equation of state for the inference via framework 2. The left panel of figure \ref{fig:9} shows the dependence of moment of inertia as a function of $\kappa$ for the BSR4 and two other equations of state (SLY4 \citep{Douchin:2001sv}, APR \citep{Akmal:1998cf}). This dependency is calculated by varying the central density to match the normalised moment of inertia for each $\kappa$ value. Once the central density is determined, we readily obtain other parameters like the mass and radius of the star. The right panel of figure \ref{fig:9} shows the dependence of the distance to the star as a function of $\kappa$. This indicates that the parameters inferred via Framework 2 are significantly influenced by the choice of the equation of state (EOS). However, the relative errors appear to be independent of the EOS (Fig \ref{fig:10}). We discuss this dependence on the equation of state in section \ref{Results}.

\subsubsection{Output parameters}\label{Output parameters:case2}

Here again we convert $\tilde{\alpha}_{\rm out}$ to  $\alpha_{s,{\rm out}}$ by using Eq \eqref{alpha}. However, unlike the previous framework, the values of $M,R$ and $\tilde{J}$ that are required in \eqref{alpha}, are calculated using $\kappa_{\rm in}$ and the choice of the equation of state (BSR4) as discussed above. 
Similar to section \ref{monte1}, we ignore data that produces $ \alpha_{s,{\rm out}} \not \in [10^{-8}, 10^{-1}]$  and $ n_{\rm out} \not\in [3,7]$. It again leads to omission of around 8\% to 16\% of data as $n_{\rm in}$ increases from $3$ to $7$. We again use the median relative error (Eq \eqref{median_error}) to compare the input and output parameters, but we don't normalise the error with the 0.1\% error assumed for the rotational frequency. This is because the error depends on other input parameters, as $T \to \infty$:
\begin{eqnarray}
\lim_{T \to \infty} \frac{\sigma(\bar{I})}{\bar{I}} &=& \Big[\Big( \frac{1}{C^2} \big(0.831 C^{-1} + 0.420 C^{-2} + 9.525 \times 10^{-3} C^{-3} \nonumber \\
&&- 1.087 \times 10^{-3} C^{-4} \big)^2 \nonumber \times \frac{f^2}{f_{\text{rot}}^2} 
\Bigg(\frac{\sigma(f_{\rm{rot}})}{(0.478 + 2.22 C)}\Bigg)^2 \\
&& + C^2 \sigma_{\kappa - C}^2\Big) + \sigma_{\bar{I} - C}^2\Big]^{1/2}. 
\end{eqnarray}

\section{Results} \label{Results}dha
\begin{figure}
    \centering
    \includegraphics[width=0.5\textwidth]{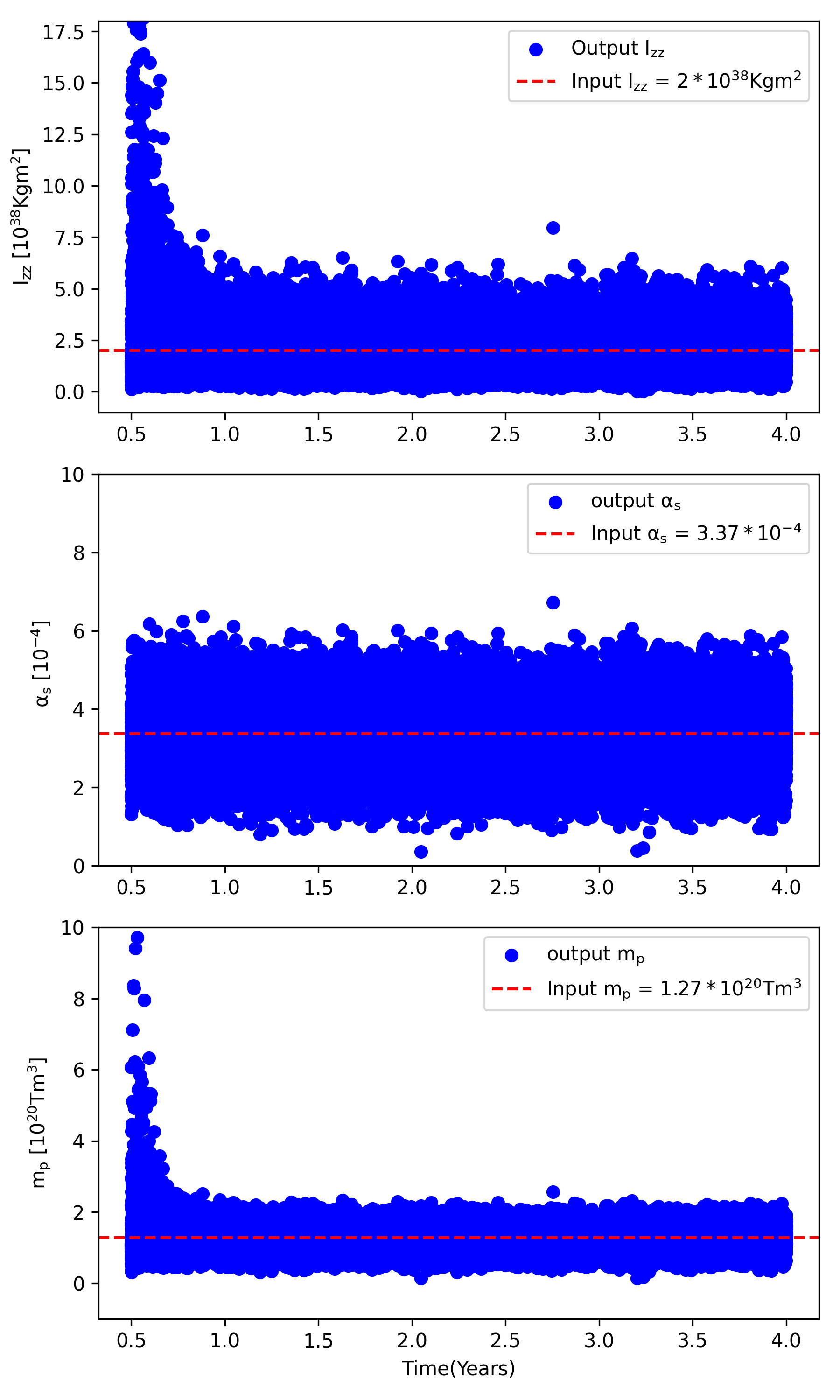}
    \caption{ Inference via framework 1: ($I_{{zz}_{\rm out}},{{\alpha_{s}}_{\rm out}}, {m_p}_{\rm out}$) converges to ($I_{{zz}_{\rm in}},{{\alpha_{s}}_{\rm in}}, {m_p}_{\rm in}$) with increase in observation time.  Here the input parameters are $f = 300$ Hz , $\dot{f} = - 10^{-9}$ Hz s$^{-1}$, $n = 3.1$, $\mathcal{D} = 30$ Hz$^{-1/2}$, $\cos(\iota) = 0$ and $I_{zz} = 2 \times 10^{38}$ kg-m$^2$. We convert the value of $\tilde{\alpha}$ into fiducial estimates of $\alpha_s$ by setting $M = 1.4 M_\odot$, $\Tilde{J} = 0.01635  $ and $R = \sqrt{5I_{zz}/2M}$. }
    \label{fig:2}
\end{figure}
In this section, we present the results of the Monte Carlo error estimation study mentioned in section \ref{Monte Carlo Simulation}. We analyse the dependence of errors on various factors like observation time and braking index and, also present a comparative analysis between both frameworks. Finally, we address the equation of state dependence of framework 2.

\begin{figure}
    \centering
    \includegraphics[width=\columnwidth]{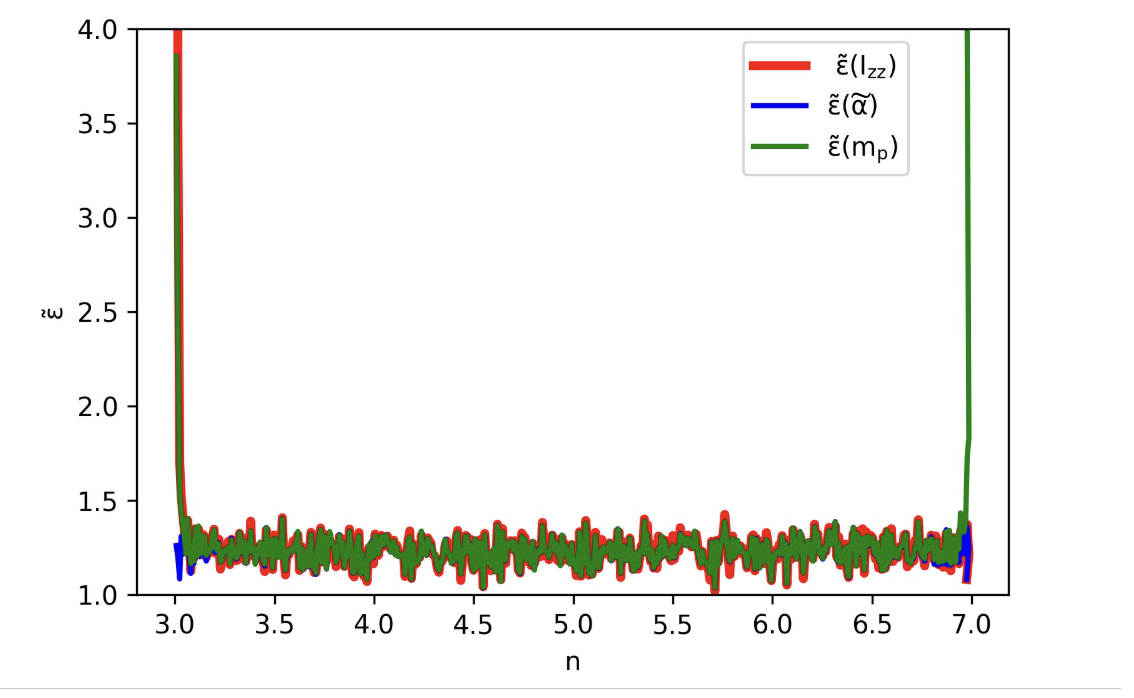}
    \caption{Inference via framework 1: Normalised relative errors ($\Tilde{\epsilon}$) as a function of the braking index $n$ post-down-sampling.  Here the input parameters are $f = 300$ Hz , $\dot{f} = - 10^{-9}$ Hz s$^{-1}$, $T = 1$ year, $\mathcal{D} = 30$ Hz$^{-1/2}$, $\cos(\iota) = 0$ and $I_{zz} = 2 \times 10^{38}$ Kg-m$^2$. }
    \label{fig:3}
\end{figure}
 The errors inferred via the first framework are dominated by the error in distance for even an observation time of $T=0.5$ years. This feature is seen for all values of the braking index except close to the extremes ($3$ or $7$). Figure \ref{fig:2} shows how the neutron star properties inferred via the first framework, converge to their actual values as observation time increases for $n = 3.1$. The input values for the parameters are $I_{zz} = 2 \times 10^{38}$ Kg-m$^2$ , $\tilde{\alpha} = 3.69 \times 10^{37}$ ( converted into fiducial estimates: $\alpha_{s} = 3.37 \times 10^{-4}$) and $m_p = 1.27 \times 10^{20}$ Tesla-m$^2$. As expected from theoretical considerations (Eqs \eqref{al_error} - \eqref{m_inf}), the errors in the parameters ($I_{zz}, m_p$) decrease with observation time and for $T>1$ years we see the error saturates to the error in the distance measurement.
Unlike the errors of other parameters, the error of $\tilde{\alpha}$ does not blow up at $n\approx3$, since it is independent of the braking index (See Eqs. \eqref{al_error}). The distance measurement dominates this error even for an observation time of $T = 0.5$ years. We  can check this  by substituting all the assumed input parameters in Eq \eqref{al_error}:
\begin{eqnarray}
\frac{\sigma(\alpha)^2}{\alpha^2} &=& \frac{\sigma(r)^{2}}{r^2} + \frac{\sigma(h_0)^2}{h_{0}^{2}} + \frac{16875 \mathcal{D}^2}{16 \pi^2 f^2 T^3} \nonumber \\
&=& 0.04 + 0.0025 + 2.75 \times 10^{-22}\, ,
\end{eqnarray}
where the observation time is $T = 0.5$ years.

Figure \ref{fig:3} shows the dependence of normalised relative errors ($\Tilde{\epsilon}$)   on the neutron star's braking index ($n$), where signals are inferred via the first framework. It is for signals with $f_{\rm in} = 300$ Hz, $\dot{f}_{\rm in} = - 10^{-9}$ Hz s$^{-1}$, $T_{\rm in}=1$ year, $\mathcal{D}_{\rm in} = 30$ Hz$^{-1/2}$,  and $I_{{zz}_{\rm in}}$ = $2 \times 10^{38}$ Kg-m$^2$. The moment of inertia ($I_{zz}$) is best estimated at ($n \approx 7$), where the spin-down is caused mostly due to gravitational waves. The perpendicular component of the dipole moment is estimated with the best accuracy when the spin-down is caused by both electromagnetic radiation and gravitational waves.  As expected, the median error in the parameter alpha ($\tilde{\alpha}$)  does not depend on the braking index.

\begin{figure}
    \centering
    \includegraphics[width=\columnwidth]{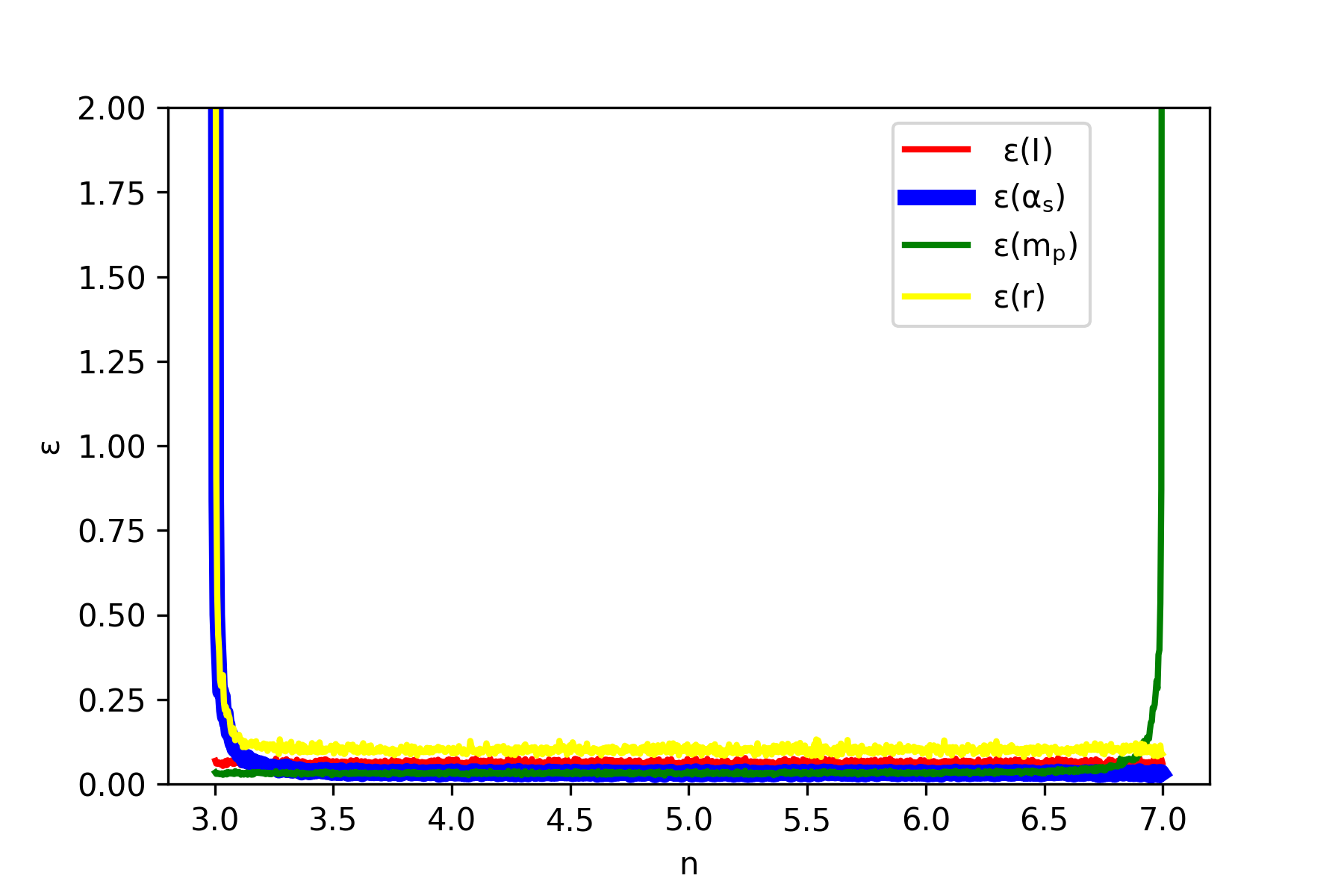}
    \caption{Inference via framework 2 : Relative errors ($\epsilon$) as a function of braking index $n$ post-down-sampling.  Here the input parameters are $h_0 = 1 \times 10^{-26}$, $f = 300$ Hz , $\dot{f} = - 10^{-9}$ Hz s$^{-1}$, $T = 1$ year, $\mathcal{D} = 100$ Hz$^{-1/2}$, $\cos(\iota) = 0$ and $\kappa = 0.56$.}
    \label{fig:4}
\end{figure}
Figure \ref{fig:4} shows the sharp contrast in the dependence of median errors ($\epsilon$) on the braking index ($n$) when we infer parameters via the second framework. This is shown for signals with ${h_0}_{\rm in} = 1 \times 10^{-26}$, $f_{\rm in} = 300$ Hz, $\dot{f}_{\rm in} = -10^{-9}$ Hz s$^{-1}$ , $T_{\rm in} = 1$ year, $\mathcal{D}_{\rm in} = 100$ Hz$^{-1/2}$, and $\kappa_{\rm in} = 0.56$ (which implies $I_{{zz}_{\rm in}} \approx 2 \times 10^{38}$ Kg-m$^2$). In this framework, the moment of inertia is independent of the braking index as it is directly inferred from the frequency of the signal. The distance of the star ($r$) and the parameter $\tilde{\alpha}$ is best estimated at $n \approx 7$, and the perpendicular component of the dipole moment can be best estimated at around $n \approx 3$. We also observe the median error in distance saturates at a greater value than other parameters as it is the only parameter that directly depends on the strain amplitude ($h_0$). For inference via the second framework, the median errors in $\tilde{\alpha}$ can be interpreted as median errors in $\alpha_{s}$, as one can directly estimate ($\alpha_{s}$) in this framework (Section \ref{Output parameters:case2}).

\begin{figure}
    \centering
    \includegraphics[width=\columnwidth]{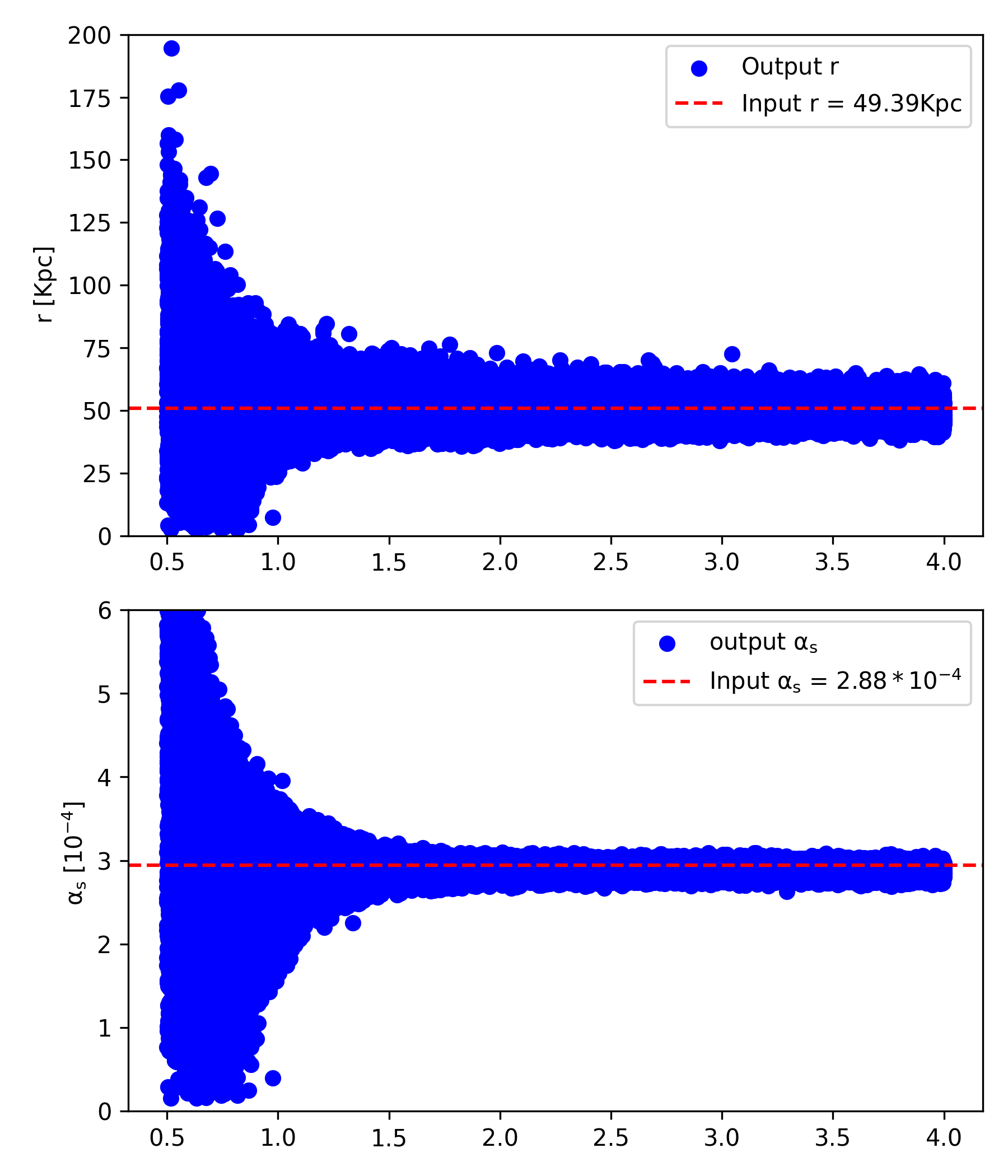}
    \caption{Inference via framework 2: ($r_{\rm out},\alpha_{\rm out}$) converges to  ($r_{\rm out},\alpha_{\rm out}$). Here the inputs are ${h_0}_{\rm in} = 1 \times 10^{-26}$, $f = 300$ Hz , $\dot{f} = - 10^{-9}$ Hz s$^{-1}$, $n = 3.1$, $\mathcal{D} = 100$ Hz$^{-1/2}$, $\cos(\iota) = 0$ and $\kappa = 0.56$. }
    \label{fig:5}
\end{figure}
In figure \ref{fig:5}, we show the convergence of ($r_{\rm out},{\alpha_s}_{\rm out}$) to  ($r_{\rm in},{\alpha_s}_{\rm in}$) for signals with $h_0 = 1 \times 10^{-26}$,  $f_{\rm in} = 300$ Hz , $\dot{f}_{\rm in} = - 10^{-9}$ Hz s$^{-1}$, $n_{\rm in}$ = 3.1, $\mathcal{D}_{\rm in} = 100$ Hz$^{-1/2}$, $\cos(\iota)_{\rm in} = 0$ and $\kappa_{\rm in} = 0.56$. We observe the expected convergence of the parameters with the increase in observation time. The parameters saturate at an observation time of $T> 2$ years due to the 0.1\% error in the rotation frequency, and $2.2\%$ (\cite{Ghosh:2022ysj}, private communication) and $9.4\%$  from $\sigma_{\kappa - C}$ and $\sigma_{\bar{I} - C}$ respectively. In comparison to $\alpha_s$, the saturation of the error in the distance measurement is dominated by the error in $h_0$. 

\begin{figure*}
    \begin{tabular}{c}
     \includegraphics[width=\textwidth]{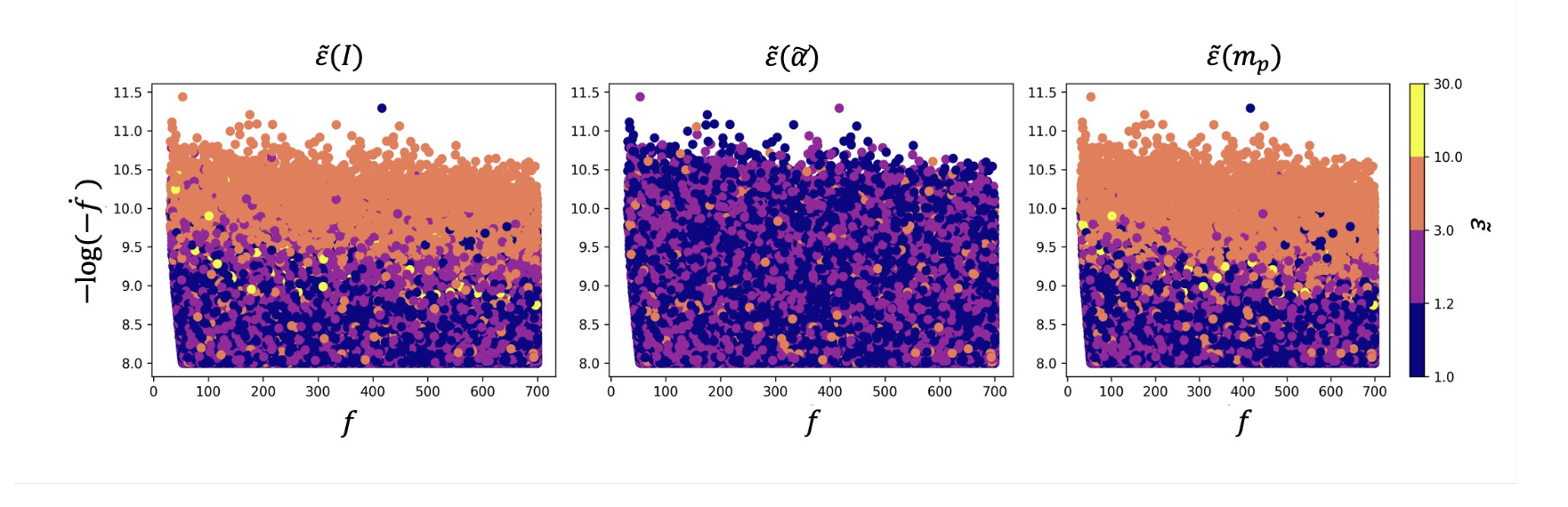} \\
     (a) $n=3.01$\\
      \includegraphics[width=\textwidth]{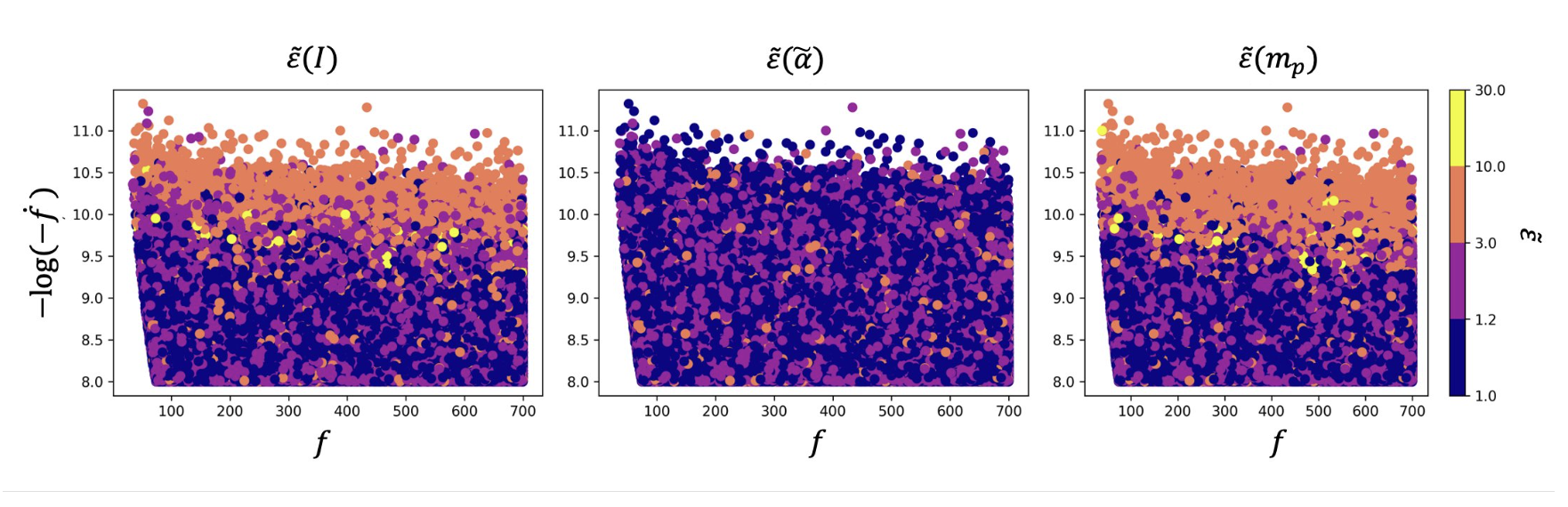}\\
    (b) $n=3.1$ \\
      \includegraphics[width=\textwidth]{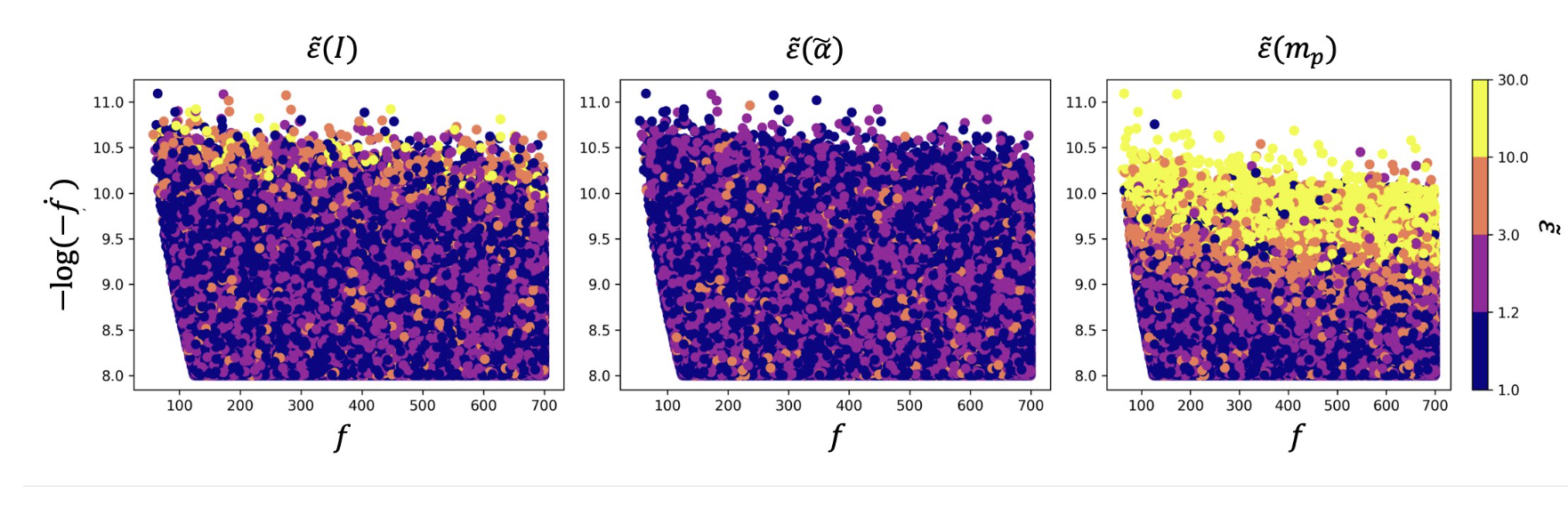}\\
    (c) $n=6.99$ 
    \end{tabular}
    \caption{Inference via framework 1: normalised relative errors $\Tilde{\epsilon}(I_{zz})$, $\Tilde{\epsilon}(\tilde{\alpha})$,  $\Tilde{\epsilon}(m_p)$, for braking index $n = 3.01,\,3.1,\,6.99$, as function of frequency ($f$) and it's derivative ($\dot{f}$). Plotted are median errors for $T = 1$ year and $\mathcal{D} = 30$ Hz$^{-1/2}$. The white areas indicate regions where $ \Tilde{\epsilon} > 30$ or $n_{\rm out} \not \in [3,7]$ or $\alpha_s \not \in [10^{-6} , 10^{-1}]$. We used a total of $10^6$ points.}
    \label{fig:6}
\end{figure*}
Figure \ref{fig:6} shows the normalised relative errors $\Tilde{\epsilon}(I_{zz})$, $\Tilde{\epsilon}(\tilde{\alpha})$, $\Tilde{\epsilon}(m_p)$ as function of frequency ($f$) and it's derivative ($\dot{f}$), for three values of braking index. The median errors are for parameters inferred via the first framework and have been taken over the sampled region of $I_{zz}$ and $\cos(\iota)$ mentioned in section \ref{case 1:Choice of input parameters input Parameters}. These errors are for signals with an observation time of  $T= 1$ year and detected with $\mathcal{D} = 30 Hz^{-1/2}$, which is relevant for a signal detected in an all-sky search.

\begin{figure*}
   \begin{tabular}{c}
    \includegraphics[width=\textwidth]{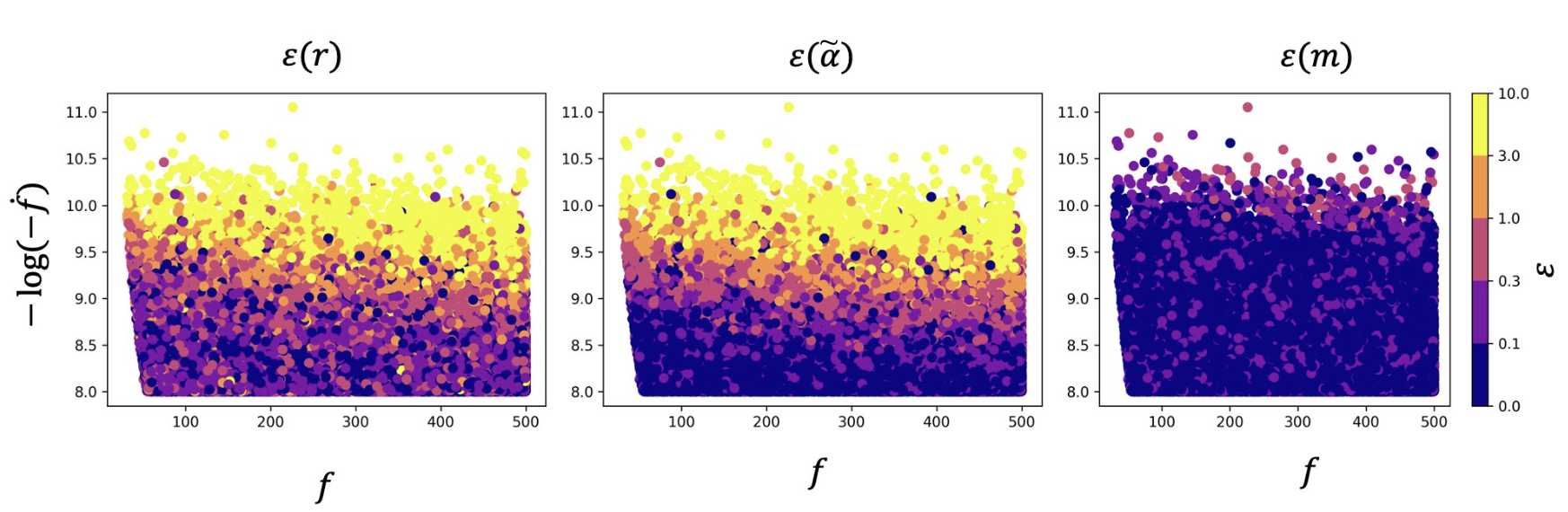}      \\
    (a) $n=3.01$ \\
     \includegraphics[width=\textwidth]{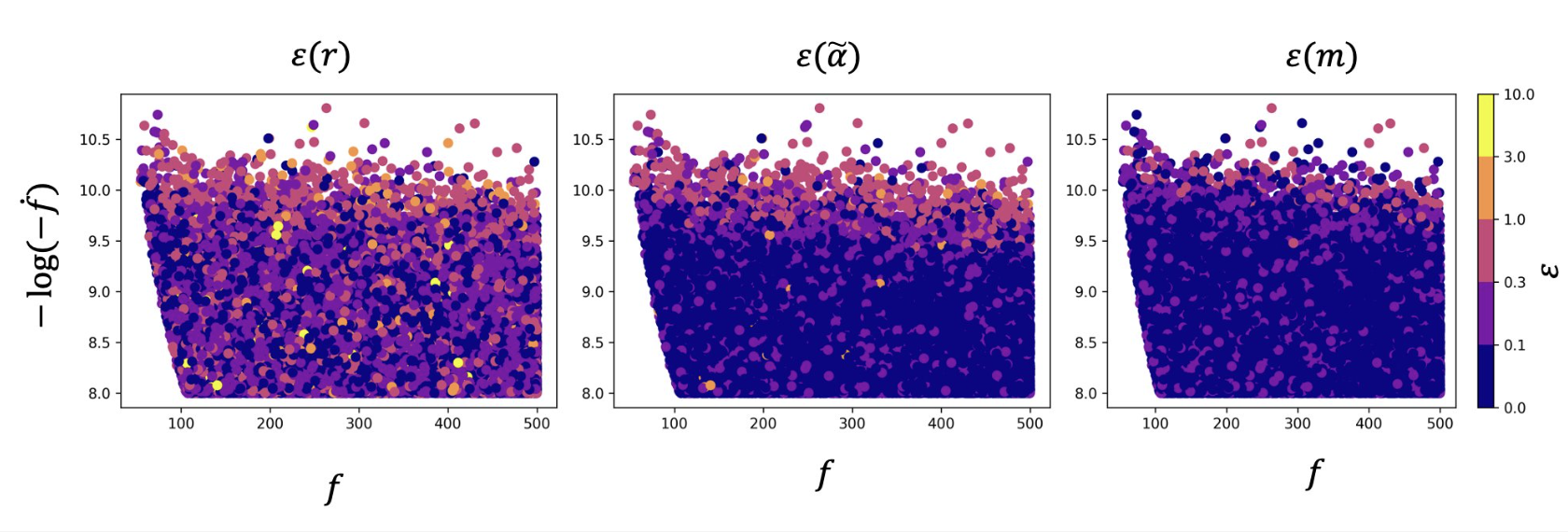}\\
     (b) $n=4.0$ \\
     \includegraphics[width=\textwidth]{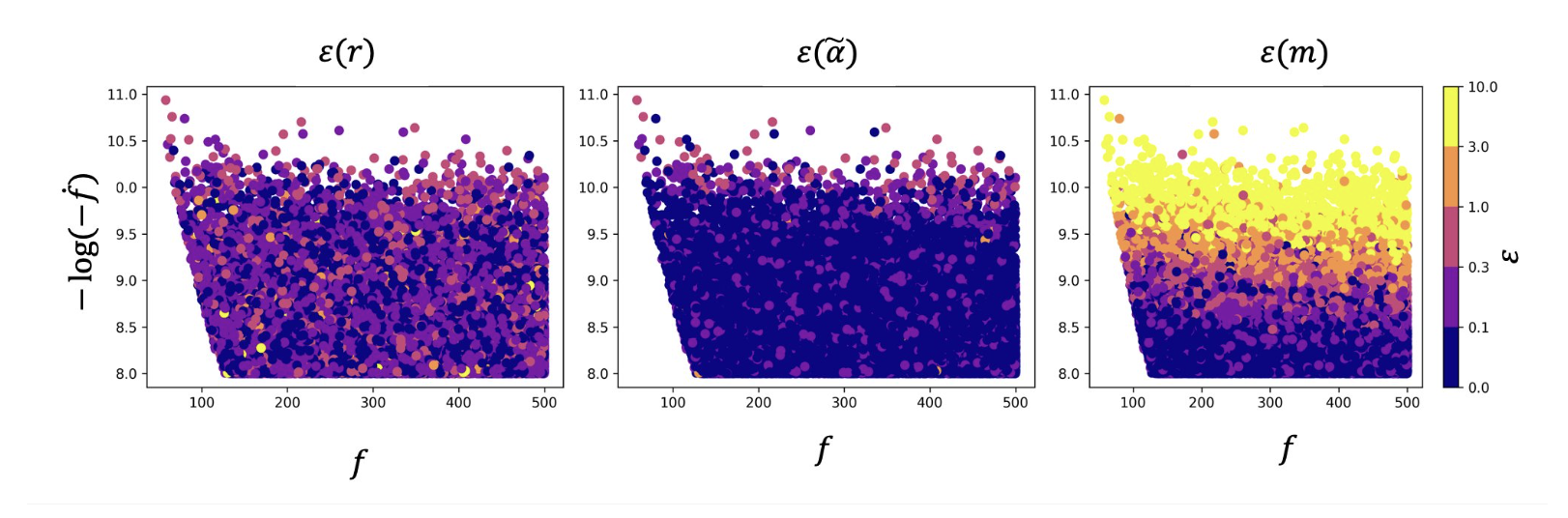}\\
     (c) $n=6.99$
   \end{tabular}
   \caption{Inference via framework 2: Relative errors $\Tilde{\epsilon}(I_{zz})$, $\Tilde{\epsilon}(\tilde{\alpha})$, $\Tilde{\epsilon}(m_p)$, for braking index $n = 3.01,4,6.99$, as function of frequency ($f$) and it's derivative ($\dot{f}$). Plotted are median errors for $T = 1$ year and $\mathcal{D} = 100$ Hz$^{-1/2}$. The white areas indicate regions where $\epsilon > 10$ or $n_{out} \not \in [3,7]$  or $\alpha_s \not \in [10^{-6} , 10^{-1}]$. Here we have used $10^5$ points.}
    \label{fig:7}
\end{figure*}
In figure \ref{fig:7} we plot the  relative errors $\epsilon(r)$, $\epsilon(\tilde{\alpha})$, $\epsilon(m_p)$ ($\epsilon(I_{zz})$ is not shown) as function of braking index ($n$), frequency ($f$) and it's derivative ($\dot{f}$), for parameters inferred via the second framework. As mentioned earlier  $\epsilon(\tilde{\alpha}) \equiv \epsilon(\alpha_s)$. These median errors are also taken over the sampled region of $h_0$ and $\cos(\iota)$ mentioned in section \ref{Choice of input parameters input Parameters:case2}. These errors are for signals with an observation time of   $T= 1$ year and detected with $\mathcal{D} = 100$ Hz$^{-1/2}$, which is relevant for a signal detected in a narrow band search.

Both fig \ref{fig:6} and fig \ref{fig:7} show that the errors in all inferred properties are minimum for high spin-down rates ($\dot{f} \approx - 10^{-8}$ Hz s$^{-1}$ and low frequency ($f \approx 100$ Hz ). For stars which are spinning down slowly ($\dot{f} < - 10^{-11}$), the errors are so high that the condition $3<n_{\rm out}<7$ cannot be satisfied. These features that are common among both frameworks are also the case for signals produced due to mountains on a neutron star and when an inference strategy similar to the first framework is used \citep{Lu:2022oys}. In addition, we identify a white region characterized by low frequencies ($f < 100 Hz$) and high spin-down rates ($\dot{f} > -10^{-9}$ Hz s$^{-1}$), whose size is minimal for $n \approx 3$ but gets larger as $n \approx 7$. This region corresponds to instances where the condition $[{\alpha_s}_{\text{in}} < 10^{-1}]$ is not met.

\begin{figure*}
 \centering
 \begin{tabular}{ccc}
  \includegraphics[width=0.31\textwidth]{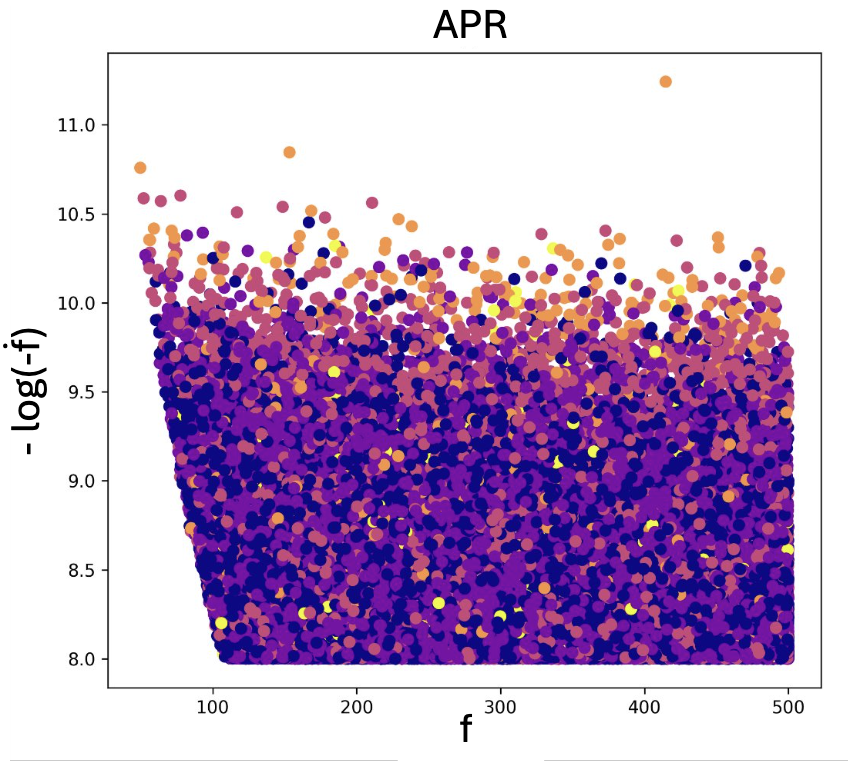} &
  \includegraphics[width=0.31\textwidth]{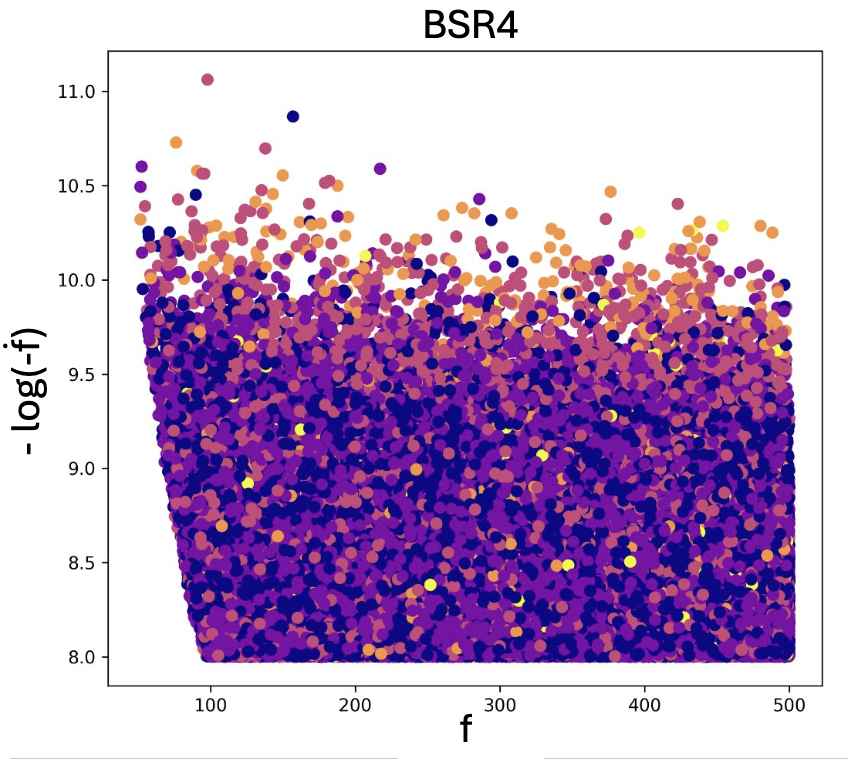} &
   \includegraphics[width=0.38\textwidth]{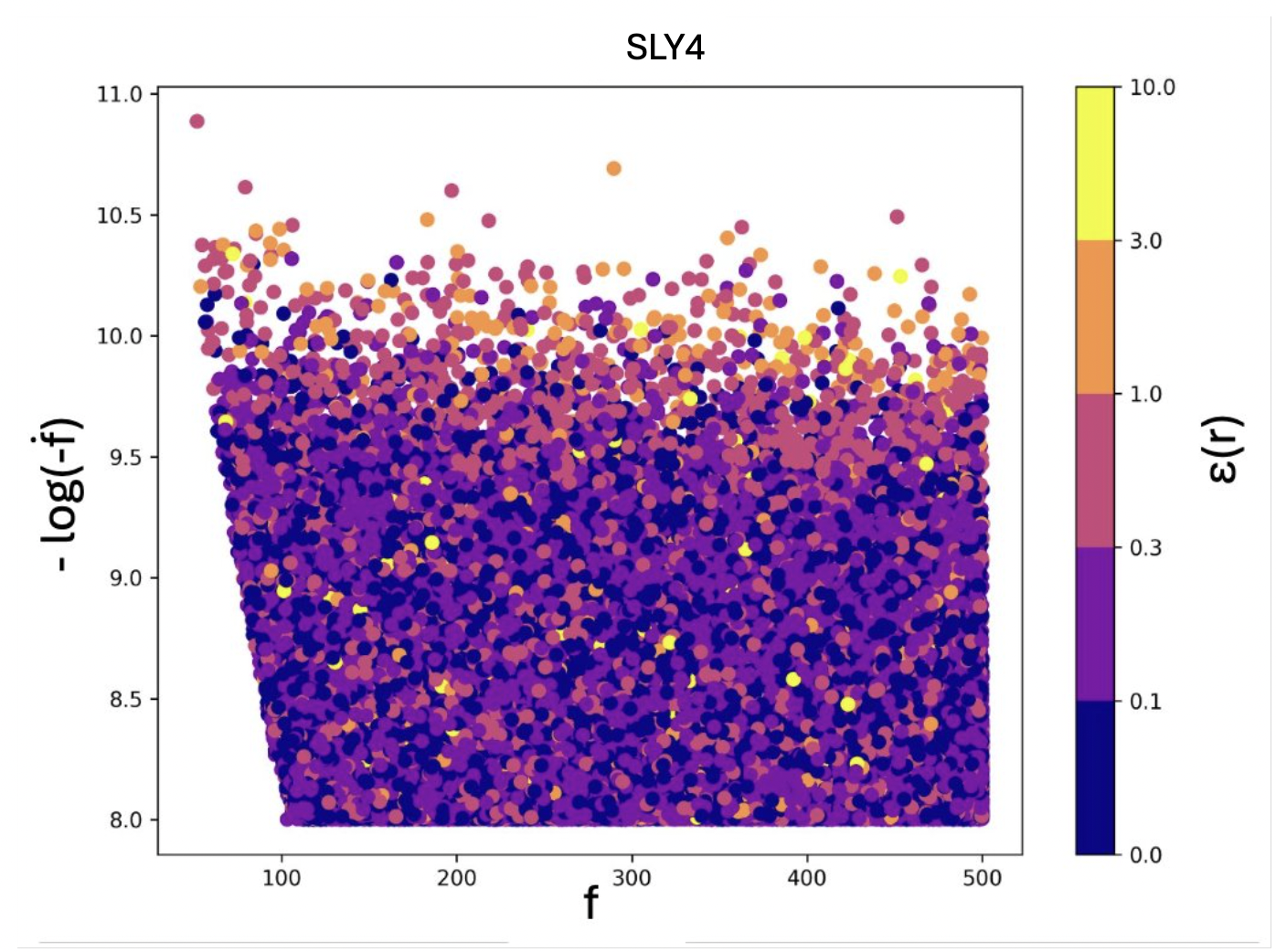}
 \end{tabular}
\caption{Relative error $\epsilon(r)$, as function of frequency ($f$) and it's derivative ($\dot{f}$), for different EOS. Plotted are errors for $n = 3.5$, $T = 1$ year and $\mathcal{D} = 100$ Hz$^{-1/2}$. The white areas indicate regions where $\epsilon > 10$ or $n_{out} \not \in [3,7]$  or $\alpha_s \not \in [10^{-6} , 10^{-1}]$. This EOS independence of $\epsilon$ is seen throughout the parameter space, only small example is shown here.}
\label{fig:10}
\end{figure*}

Fig \ref{fig:6} also suggests that a normalised relative error $\Tilde{\epsilon} \le 1.2$ can be achieved for most of the parameter space ($f,\dot{f}$) for all sky searches. This implies (via Eqn \eqref{normal}) an error of 32\% in $I_{zz}$ and a 16\% error in $\tilde{\alpha}$ and $m_p$. The errors in $\tilde{\alpha}$ are sufficiently small that converting the measured $\tilde{\alpha}$ into fiducial estimates of $\alpha_s$ (as mentioned in section \ref{Output Parameters:case 1}) would provide valuable insights into the damping mechanisms that limit the growth of the r-modes \citep{Arras:2002dw, Bondarescu:2008qx, Bondarescu:2007jw, Alford:2012yn}. On the other hand, Fig \ref{fig:7} shows that framework 2 leads to much lower errors due to three key factors: the high accuracy in pulsar frequency measurements, the use of universal relations and the fact that it does not depend on electromagnetic distance measurements. An error of 32\% can be achieved in distance measurements, which is comparable to electromagnetic observations \citep{Dis_error}. We also note a sufficiently low error of at least $20\%$ for $I_{zz}$ and at least $10\%$ for both $\tilde{\alpha}$ and $m_p$ (refer to Fig. \ref{fig:7}), achievable with a sufficiently high spin-down rate. This is because the errors displayed in all properties, except the distance, do not directly depend on the strength $h_0$ of the detected signal (as we have assumed $\mathcal{D} = 100$ Hz$^{\frac{1}{2}}$). However, this approach comes with the drawback of assuming a specific equation of state and works only for slowly rotating neutron stars. Although Figure \ref{fig:9} shows a strong dependence on the EOS, the relative errors appear to be independent of the EOS (Fig \ref{fig:10}). A similar conclusion was obtained in Ref. \citep{Ghosh:2023vja}.

Considering the significant dependence of the inferred parameters on the EOS, it’s reasonable to question whether this inference holds any practical value. However, we believe they remain valuable, as a broad range of realistic EOS models can be used to establish robust bounds on these parameters—particularly on the distance to the star (Figure \ref{fig:9}). Secondly, If the detected CGW signal is sourced from a star whose accurate distance measurement exists from other electromagnetic sources (e.g. Crab Pulsar), then Framework 2 can be utilized to constrain the EOS of the Star.
Finally, Framework 2 has the potential to be independent of the EOS when mass measurements through electromagnetic observation are available. In particular, Low-mass X-ray binaries (LXMBs) represent crucial candidates for signals generated by r-modes or their relativistic counterparts \citep{Kokkotas:2015gea}. Accurate mass measurements through electromagnetic observations are possible for such systems.  Consequently, if we detect a CGW, and such mass measurements exist, one can directly substitute it in the $I-C$ relation to estimate $I_{zz}$, $m_p$, and $r$ without assuming any specific equation of state.

\section{Discussion and Conclusion} \label{dis}

In this article, we present an analysis similar to \citep{Lu:2022oys} of what properties can be inferred from neutron stars that radiate electromagnetic waves and detectable continuous gravitational waves produced by r-modes and their relativistic equivalents (ALH modes). We investigate two different frameworks. In the first framework, we assume the distance of the star is measured via electromagnetic observations with 20\% accuracy. We then infer three neutron star properties: its principal moment of inertia ($I_{zz}$), the component of magnetic dipole moment perpendicular to the rotation axis ($m_p$), and a parameter ($\tilde{\alpha}$) which is related to the saturation amplitude by $\tilde{\alpha} = \alpha_{s} M R^3 \tilde{J}$.  Unlike the signals produced due to mountains, the signals due to ALH-mode oscillations detected via narrow-band searches give us information on an additional parameter ($\kappa$) which satisfies universal relations with the compactness of the star \citep{Ghosh:2022ysj, Idrisy:2014qca}. In the second framework, we use this and the $I-C$ universal relations to directly measure the distance ($r$) of the neutron star, along with the three parameters mentioned above.   We then use a simple Fisher information matrix-based approach to present a quantitative error estimation study for parameters inferred via both frameworks.

Monte Carlo simulations typical for all-sky searches are done for the first framework, whereas simulations typical for narrow-band searches are done for the second framework. When inferring properties via the first framework, for detected signals with a year-long observation time (which could be higher depending on the detector duty cycle), it is possible to achieve an accuracy of 32\%  for $I_{zz}$ and 16\%  for $\tilde{\alpha}$ and $m_p$. On the other hand, for inference via the second framework under similar conditions, we observe a comparatively higher accuracy of $10-20\%$ or less is achievable for $I_{zz}$, $\tilde{\alpha}$, and $m_p$, and 32\% accuracy for $r$. The most accurate estimates are when the braking index is in the region $n \in [4,6]$, $f$ is small, and $\dot{f}$ is large.

We note that we can't directly extract $\alpha_s$ but only the parameter $\tilde{\alpha}$ ($\tilde{\alpha} = \alpha_s MR^3\tilde{J}$) using the first framework. Even if mass and radius measurements become available from electromagnetic observations, there remains an equation of state-dependent parameter $\tilde{J}$ necessary for estimating the saturation amplitude ($\alpha_s$). It was demonstrated  \citep{Alford:2012yn} that this parameter is bounded within a factor of 2, suggesting that a fiducial value of $\tilde{J} \approx 0.01635$ could provide a reasonable rough estimate. A key drawback of the second framework is that it requires an assumption of a neutron star equation of state and works only in a restricted parameter space. It is noteworthy that while the values of the parameters change with the equation of state, the errors remain largely consistent. Despite EOS dependence, robust bounds on parameters like distance can be derived using realistic EOS models. For stars with known distances (e.g., Crab Pulsar), Framework 2 could constrain the EOS, and with precise mass measurements (e.g., in LXMBs that are key candidate sources for r-modes or ALH modes), it can estimate $I_{zz}$, $m_p$, and $r$ independently of the EOS.

We plan to perform parameter inference by adopting a Bayesian framework in the future, as it would yield robust conclusions and incorporate priors from electromagnetic observations of neutron stars. One could also explore alternate spin-down models encompassing current braking index observations  \citep{spin-down_review} and study more complex magnetic field models rather than a simple dipolar magnetic field to expand on this work \citep{Complex_mag}.

\section*{Acknowledgments}
We would like to thank Dr. Suprovo Ghosh for assisting us with the fit error of one of the universal relations we used.  

\section*{Data Availability}
The data generated in this work are not publicly available but are available from the corresponding author upon reasonable request.

\bibliographystyle{mnras}

\bibliography{References}

\bsp	
\label{lastpage}
\end{document}